\documentclass[12pt,preprint]{aastex}

\shorttitle{[OIII] images of Seyferts}
\shortauthors{Schmitt et al.}
\begin{document}

\title{Multiwavelength star formation indicators: Observations\altaffilmark{1,2}}

\author{H. R. Schmitt\altaffilmark{3,4,5,6,7}, D. Calzetti\altaffilmark{8},
L. Armus\altaffilmark{9}, M. Giavalisco\altaffilmark{8},
T. M. Heckman\altaffilmark{8,10}, R. C. Kennicutt
Jr.\altaffilmark{11,12}, C. Leitherer\altaffilmark{8},
and G. R. Meurer\altaffilmark{10}}

\altaffiltext{1}{Based on observations made with the NASA/ESA Hubble Space
Telescope, which is operated by the Association of Universities for Research
in Astronomy, Inc., under NASA contract NAS5-26555.}
\altaffiltext{2}{Based on observations obtained with the Apache Point Observatory
3.5-meter telescope, which is owned and operated by the Astrophysical Research
Consortium.}
\altaffiltext{3}{Remote Sensing Division, Code 7210, Naval Research Laboratory, 4555 Overlook Avenue, Washington, DC 20375}
\altaffiltext{4}{Interferometrics, Inc., 13454 Sunrise Valley Drive, Suite 240, Herndon, VA\,20171}
\altaffiltext{5}{email:hschmitt@ccs.nrl.navy.mil}
\altaffiltext{6}{Visiting Astronomer, Cerro Tololo Inter-American Observatory,
National Optical Astronomy Observatories, which are operated by AURA, Inc.,
under a cooperative agreement with the National Science Foundation}
\altaffiltext{7}{Visiting Astronomer Kitt Peak National Observatory,
National Optical Astronomy Observatories, which are operated by AURA, Inc.,
under a cooperative agreement with the National Science Foundation}
\altaffiltext{8}{Space Telescope Science Institute, 3700 San Martin Drive, Baltimore, MD21218}
\altaffiltext{9}{Spitzer Science Center, California Institute of Technology, Mail Stop 220-6,
Pasadena, CA 91125}
\altaffiltext{10}{Department of Physics and Astronomy,
Johns Hopkins University, Baltimore, MD21218}
\altaffiltext{11}{Steward Observatory, University of Arizona, 933 North Cherry
Avenue, Tucson, AZ 85721}
\altaffiltext{12}{Institute of Astronomy, University of Cambridge, Madingley
Road, Cambridge, CB3 0HA,   UK}

\begin{abstract}
We present a compilation of multiwavelength data on different star formation
indicators for a sample of nearby star forming galaxies. Here we discuss the
observations, reductions and measurements of ultraviolet images obtained
with STIS, on board the Hubble Space Telescope ($HST$), ground-based
H$\alpha$, and VLA 8.46~GHz radio images. These observations are complemented
with infrared fluxes, as well as large aperture optical radio and ultraviolet
data from the literature. This database will be used in a forthcoming paper to
compare star formation rates at different wavebands. We also present spectral
energy distributions (SEDs) for those galaxies with at least one far-infrared
measurements from ISO, longward of 100$\mu$m. These SEDs are divided in two
groups, those which are dominated by the far-infrared emission, and those
where the contribution from the far-infrared and optical emission is
comparable. These SEDs are useful tools to study the properties of high
redshift galaxies.
\end{abstract}

\keywords{galaxies: evolution -- galaxies: starburst -- stars: formation --
infrared: galaxies -- radio continuum: galaxies -- ultraviolet: galaxies}

\section{Introduction}
The global star formation history of the Universe is one of the
crucial ingredients for understanding the evolution of galaxies as a
whole, and for discriminating among different formation scenarios,
(Madau, Pozzetti \& Dickinson 1998; Ortolani et al. 1995; Baugh et al.
1998). Star formation is traced by a number of observables, often
complementary to each other. The ultraviolet (UV) is where the bulk of
the energy from young, massive stars is emitted, but it suffers from
the effects of dust extinction. The nebular emission line H$\alpha$ is
a tracer of ionizing photons, and thus of young, massive stars, but is
also affected by assumptions on the stellar initial mass function. The
far-infrared (FIR) emission comes from the dust-processed stellar
light and reliably tracks star formation in moderately to very dusty
galaxies, but is not a narrow band measurement (like the UV or
H$\alpha$) and needs to be measured across the entire infrared
wavelength range. Finally, the non--thermal radio emission is a tracer
of the supernova activity in a galaxy, but suffers from uncertain
calibration.

In recent years, all wavelength regions have been exploited to track
down star formation at all redshifts, and trace the star formation
history of the Universe. At high redshift the Lyman-break galaxies
(Steidel et al. 1996, 1999) provide the bulk of the UV light from star
formation, while SCUBA has been used to detect the brightest FIR sources
in the range 1$\lesssim$z$\lesssim$3--4 (e.g., Barger et al. 1999,2000;
Smail et al. 2000; Chapman et al. 2003). The Lyman-break galaxies
are, by selection, actively star-forming systems, resembling in many
aspects local starburst galaxies (Pettini et al. 1998; Meurer et al. 1997;
Meurer, Heckman \& Calzetti 1999) , and possibly covering a large range in
metal and dust contents (Pettini et al. 1999; Steidel et al. 1999;
Calzetti 1997). The SCUBA sources occupy the high-end of the FIR luminosity
function of galaxies (Blain et al. 1999), and are dust-rich objects. The
relationship between the UV-selected and FIR-selected luminous systems is
not yet clear. Whether the
two types of galaxies represent two distinct populations or the two
ends of the same population is still subject to debate (Adelberger \&
Steidel 2000). The answer to this question can drastically change our
understanding of the evolution, star formation history and metal/dust
enrichment of galaxies.

The difficulty in relating the two types of galaxies stems from the
paucity of {\it simultaneous} UV, H$\alpha$, FIR($>$100~$\mu$m), and
radio data for both low- and high-z galaxies. While multiwavelength
observations of high-z galaxies are presently hampered by the
sensitivity and positional accuracy limitations of current
instrumentation, multiwavelength observations of local galaxies are
within the accessible realm and are key for producing the spectral
energy distributions (SEDs) that can be used as templates for higher
redshift observations. To be representative, such SEDs should cover as
completely as possible the parameter range of properties (metallicity,
luminosity, etc.) of galaxies.

This paper presents a large, homogeneous dataset of multiwavelength
observations of 41 local (closer than 120~Mpc) star--forming
galaxies, spanning from the UV to the radio. The UV observations are
from our own HST/STIS imaging centered at 1600~\AA. The optical data
(H$\alpha$) and radio (6~cm and 21~cm) data are from ground-based
programs complementary to the HST observations. The FIR data are from
archival IRAS and ISO observations spanning the wavelength range
12-200~$\mu$m.

The data in this paper provide a uniform dataset, in terms of type of
data, reduction and calibration strategies in each wavelength range,
that will be used in an accompanying paper to re-evaluate star formation
rate indicators at the different wavelengths.

\section{Sample}

The global list of galaxies in our sample is presented in Table~\ref{tabcar},
where we give their coordinates, radial velocities, distances, diameters,
morphological types and foreground Galactic reddening. We gave preference
to Galactic reddening values from Burstein \& Heiles (1982) instead of
Schlegel et al. (1998), because in the case of galaxies with high foreground
reddening, like NGC\,1569, the latter values result in too high a correction
(i.e., an unphysically blue spectrum). Nevertheless, this correction
is usually very small and the difference between the two values does
not introduce a significant variation between the final fluxes.
The sample is composed of 41 galaxies, which were selected according to the
following criteria: 1-) IRAS 100~$\mu$m flux $\gtrsim$1~Jy, to ensure
detection with the ISO long wavelength camera; 2-) recession velocity
$\le$9000~km~s$^{-1}$, to resolve spatial scales as small as 35~pc with STIS,
which is the scale of large OB associations or giant molecular clouds; 3-) mean
angular diameter D$_{25}\lesssim$4$^{\prime}$, to ensure that the sources were
seen as point-like by the ISO long wavelength camera;
4-) L(H$\alpha$)$>$10$^{39}$ erg~s$^{-1}$ in the central
5$^{\prime\prime}-10^{\prime\prime}$, to select active, centrally star-forming
objects; 5-) absence of or, at worst, weak non-thermal nuclear activity, as
determined by the nuclear emission line properties of the galaxies (e.g. Ho
et al. 1997). We retained only 4 of these active galaxies in our sample, those
where the nuclear spectrum (inner $\sim$2\arcsec) shows emission line ratios
typical of AGNs, while a more extended spectrum, encompassing a region of
10\arcsec\ or larger, has emission line ratios typical of HII regions.
We estimate, based on the comparison of the nuclear and integrated H$\alpha$
or radio emission that the AGN contributes to, at most, 10\% of the total
luminosity in these galaxies.

Our sample specifically excluded ultraluminous infrared galaxies
(ULIGs), because data for a significant number of such sources are
available from the literature (Goldader et al. 2002), obtained in a
intrumental configuration similar to ours. Our project aimed at
covering the IR luminosity range of more normal galaxies, thus
extending the low bound of this parameter by a factor $\sim$10$^3$, in
order to accomodate the fact that high redshift galaxies may not all
be as IR-luminous as ULIGs. Nevertheless, we collected the data available
on ULIGs in the literature and will take them into consideration in our
analysis paper (Schmitt et al. 2005b).

The final sample covers
a wide range of intrinsic properties. The host galaxy morphologies vary
from regular spirals to interacting/merging systems. The bolometric
luminosities, calculated based on the infrared luminosities and the
correction factor given by Calzetti et al. (1999) vary by a factor of 
almost a thousand, with NGC\,1569 having L$_{BOL}\sim7\times10^8~L_{\odot}$,
while IC\,1623 has 4.7$\times10^{11}~L_{\odot}$, close to the ULIG limit.
This sample also covers a large factor in star formation activity, from the
post-starburst in NGC\,1569 up to the FIR-luminous starburst in IC\,1623;
as well as a factor of $\sim$10 in metallicity, from metal poor galaxies like
TOL\,1924-416, which has [O/H]=8.0, to metal rich ones like NGC\,7552,
which has [O/H]=9.3 (Storchi-Bergmann, Calzetti \& Kinney 1994).
Although this sample covers a wide range of intrinsic properties, one should
keep in mind that it was culled from the ISO archive, so the galaxies
may have been observed because of some characteristic of particular
interest for the original observers. As a result, the final sample may
not necessarily be representative of typical galaxies. 

Complete UV-H$\alpha$-radio coverage is achieved for 26 out of 41 galaxies,
due to various limitations (e.g. snapshot nature of the HST observations,
weather, declination of the source). Complete coverage using our own data
was obtained for 13 galaxies, with the other 13 using literature data to
supplement our own data. For an additional 11 galaxies we have data for 2
of the 3 wavelengths (UV, H$\alpha$, or radio).

\section{Observations and Reductions}

In this section we present the details of our multiwaveband observations,
reductions and measurements. We also present ultraviolet, optical, infrared
and radio data collected from the literature, which are complementary to
our observations and will be used in a forthcoming paper. Table~\ref{tabobs}
presents the details about the different observations.

\subsection{Ultraviolet Observations}

Ultraviolet images were obtained for 22 of the galaxies in our sample
($\sim$50\%), using STIS on board the HST. The observations of 21 of the
galaxies were done as part of the snapshot project 8721 (P.I. Calzetti),
using the FUV-MAMA detector and the filter F25SRF2. We also obtained HST
archival data for IC\,1623, which was observed with the same configuration
by the project 8201 (P.I. Meurer). The names of the datasets in the HST
archive and exposure times of the images are given in Table~\ref{tabobs}.
Our images have a pixel size of 0.025\arcsec\ and a field of view of
$\sim$25\arcsec$\times$25\arcsec\ (1024$\times$1024 pixels). The filter used for these
observations (F25SRF2) is centered at $\lambda$1457\AA, and has a bandwidth of
284\AA, which is slightly contaminated by geocoronal OI$\lambda$1302\AA.
Guiding problems happened during the observations of MRK\,555, which resulted
in the drifting of the spacecraft. This produced slightly elongated
point sources, but did not affect the integrated flux of this galaxy.

The reduction and calibration of the data followed standard procedures.
The images of the galaxies in our snapshot project were obtained in single
exposures of 1320 seconds each. Only IC\,1623, had its observations split
into multiple exposures, which had to be registered and combined. Background
levels and standard deviations ($\sigma$) were determined on emission free
regions of the images. The images were background subtracted, clipped at the
3$\sigma$ level (pixels with flux lower than 3$\sigma$ were set to zero)
and flux calibrated using the information available in their
headers. The UV fluxes of the galaxies, obtained integrating the emission
in the final images, as well as the 3$\sigma$ detection limits, are given
in Table~\ref{tabuvha}. These values are not corrected for Galactic extinction.

Throughout this paper we use the technique of measuring the fluxes
of the images inside the 3$\sigma$ level, and not by integrating everything
inside a given aperture. The reason for doing a cut at the 3$\sigma$ level
is to avoid spurious background variations and flatfielding errors. We tested
the effect this technique has in the fainter, lower surface brightness
sources, by comparing measurements of both H$\alpha$ and UV done in this
way, with measurements done without clipping the images at the 3$\sigma$ level.
We find that in most of the cases the difference between the two techniques
is negligible, while in the worst cases it can represent a 5\% difference
in the total H$\alpha$ flux, well within our measurement uncertainties (see
below). Another way to estimate that this approach will cause an insignificant
error to the integrated fluxes is based on the fact that the area
covered by pixels between 3 and 6$\sigma$ is small, as can be seen in the
figures, suggesting that the area associated with the object and covered
by pixels below 3$\sigma$ should also be small. Combining this information
with the fact that the 3$\sigma$ flux level is very small compared to the
regions where the bulk of the emission originates, indicates that this
approach will not change the measured flux significantly. The errors
involved in the UV flux measurements are of the order of 5\%. These errors
are mostly due to the stability of STIS and the accuracy of the flux
calibration. We estimate that the contribution from Poisson noise to the
error budget is very small, accounting to less than $\sim$1\% even in the
case of the fainter galaxies observed (NGC\,3079 and NGC\,4088). For simplicity
we assume a uniform 5\% flux error for all sources.

\subsection{H$\alpha$ Observations}

The H$\alpha$ images presented in this paper were obtained in 4 observing
runs on 3 different ground based telescopes (APO, KPNO and CTIO).
We also use archival HST narrow band images of NGC\,1569 and NGC\,4214.
These images were complemented with large aperture H$\alpha$
fluxes from the literature, available for several galaxies which we
could not observe.

The APO observations were done with the 3.5~m telescope on the
second half of the night of 2000 July 3/4. We used SPIcam, binning 2 pixels 
in both direction, which gives a scale of 0.28\arcsec pixel$^{-1}$ and a
field of view of 4.8\arcmin $\times$4.8\arcmin\ (1024$\times$1024 pixels).
Two filters were used for these observations, one centered
at 6450\AA, with a bandwidth of 100\AA, and one centered at 6590\AA, with a
bandwidth of 25\AA. These filters were used for continuum and line observations,
respectively. The continuum observations consisted
of 2 exposures of 300 seconds for NGC\,6217 and 
2 exposure of 90 seconds for NGC\,6643. The on-band exposure times are give
in Table~\ref{tabobs}. We also observed spectrophotometric
standard stars from Oke (1990).

The CTIO observations were done with the 1.5m telescopes on the nights of
2000 August 01/02 -- 08/09. We used the focal ratio f/13.5 and the detector
Tek~2k \#6, which gives a scale of 0.24\arcsec pixel$^{-1}$ and a field of
view of 8.2\arcmin $\times$8.2\arcmin\ (2048$\times$2048 pixels). The filters
and integration times used for the narrow band H$\alpha$ observations are given
in Table~\ref{tabobs}. We also obtained broad R band images for each one of the
galaxies, which were used to subtract the continuum contribution from the
line images. These broad band images usually were split into 3
exposures of 500~seconds each. We observed spectrophotometric standard
stars from Stone \& Baldwin (1983) and photometric standards from Landolt (1992)
for the calibration of the narrow and broad band images, respectively.

The KPNO observations were done with the WIYN 3.5m telescope on two different
runs, on the nights of 2001 May 16/17 -- 18/19, and 2001 November 05/06 --
06/07 (we refer to these runs as WIYN-1 and WIYN-2, respectively, in
Table~\ref{tabobs}). We used the Mini-Mosaic, which gives a pixel scale of
0.14\arcsec pixel$^{-1}$ and a field of view of 9.6\arcmin $\times$9.6\arcmin\
(4096$\times$4096 pixels). The filters and integration times used for the
narrow band H$\alpha$ observations are indicated in Table~\ref{tabobs}.
As for the CTIO observations, we obtained 2 or 3 images of
120 seconds each in the R band, which were used for the continuum subtraction.
The narrow band images were calibrated using observations of spectrophotometric
standards from Oke (1990), while the broad band ones were calibrated using
standard stars from Landolt (1992).

The data reductions followed standard IRAF procedures, which started with
the overscan and bias subtraction, and the division of the images by
normalized flat-fields. The individual images of each galaxy were aligned,
convolved to bring both the line and continuum images to a similar PSF size,
and combined to eliminate cosmic rays. The background was determined from
emission free regions around the galaxy, fitted with a
polynomial function to eliminate residual illumination gradients, and
subtracted. The data were calibrated using the standard star observations,
and the World Coordinate System, obtained from stars in the field of the
galaxy, was added to the image headers. Whenever needed the fluxes were
also corrected for redshift effects, taking into account where the H$\alpha$
line fell on the transmission curve of the filter.

The continuum subtraction was done in two different ways. In the case of
the APO observations, the continuum images were not contaminated by line
emission. In this case, these images were scaled to match the width of the
line filter, and subtracted. We confirm, by checking stars around the
galaxy that this procedure did not over or undersubtract the continuum.
In the case of all the other observations, we used a similar technique
for the continuum subtraction, but had to do it recursively. Since the
R band images, which were used for the continuum, are contaminated by line
emission (1.5\% of the integrated flux on average), they have to be
corrected for this contribution to their flux,
otherwise a simple subtraction of the scaled R band image from the line image
will underestimate the total H$\alpha$ flux. The recursive subtraction was
done by first subtracting the scaled continuum image from the line image.
The resulting line image was then scaled and subtracted from the continuum
image, to remove the emission line contribution to this image,
and the corrected continuum image was then used to subtract the
continuum emission from the original line image. This process was
repeated a few times, until the H$\alpha$ and continuum fluxes
in regions affected by contamination changed by less than 0.5\% between
consecutive iteration. This indicated that the process
have converged. We usually needed only 2 to 3 iterations to reach this level.

The noise of the final continuum free images was determined on emission free
regions of the frame, and the images were clipped at the 3$\sigma$ level.
Total H$\alpha$ fluxes [F(H$\alpha$)$_{int}$] were measured by integrating
all the emission in these images. For a few of the galaxies (e.g. NGC6753,
TOL1924$-$416) the flux measurement had to take into account foreground stars.
In most of the cases the stars lie in the outskirts of the galaxies,
in areas without any H$\alpha$ emission. The area around the stars was excluded
from the flux measurements, in order to avoid possible subtraction
residuals that cloud artificially increase the flux. The only galaxy for which
a foreground star could present a problem is TOL1924$-$416 (Figure~\ref{fig13}).
However, even in this case the error caused by the star should not account
for more than 1\% of the integrated flux, since the residuals are not very large
and the star is in a region of faint H$\alpha$ emission.

For those galaxies with H$\alpha$ and UV
observations, we rotated and registered the two images, and measured the
H$\alpha$ flux inside a region matching the one covered by the UV observations
[F(H$\alpha$)$_{match}$]. These fluxes and detection limits are given in
Table~\ref{tabuvha}, where we also give integrated H$\alpha$ fluxes obtained
from the literature, for those galaxies which we did not observe. For
NGC\,1672 and NGC\,5383, we were also able to obtain published H$\alpha$ 
fluxes in regions matching the ones observed in the UV. A comparison between
F(H$\alpha$)$_{match}$ and F(H$\alpha$)$_{int}$ shows that most of the H$\alpha$
flux originates in the region covered by the UV observations. The median
F(H$\alpha$)$_{match}$/F(H$\alpha$)$_{int}$  value for our galaxies is 0.92.

Contamination by [NII]$\lambda$6548,84\AA\ is a concern for these measurements,
since it can make a considerable contribution to the flux of the images.
We deal with this problem by using spectroscopic data from the literature,
preferentially large aperture spectra. In this way we minimize problems such as
enhanced [NII] emission at the nucleus, which can be due to higher metallicity,
or even the presence of a low luminosity AGN. We give in Table~\ref{tabuvha}
the [NII]/H$\alpha$ ratios, slit areas through which they were observed,
and the H$\alpha$ fluxes corrected for [NII]
contamination (F(H$\alpha$)$_{int}^{cor}$ and F(H$\alpha$)$_{match}^{cor}$).
This correction was calculated based on the observed [NII]/H$\alpha$ ratios,
the redshifts of the galaxies and on-band filter transmission curves used to
observe them. Notice that these corrected values are likely to be a lower limit
of the real flux, since the observed emission line ratio may still be
dominated by the brighter regions of emission.
Since these regions usually are related to the nucleus, which has higher
metallicity, they should also have higher [NII]/H$\alpha$ ratios.

Finally, we estimate the errors in our flux measurements. A significant
part of the errors come from the flux accuracy of the calibration stars,
which is of the order of 4-5\%. Other effects that can introduce errors of a few \%
in the fluxes are residuals from the flat-field correction, variations of the
sky transparency overnight, uncertainties in the Galactic foreground reddening
correction, and residuals from the sky and continuum subtraction.
Poisson noise can also introduce some uncertainties in the flux measurements,
but this source of error is much smaller than 1\% in our case.
Taking all these sources of error into account, we make the conservative
assumption that the error in our flux measurements is of the order of 10\%.

\subsection{Radio Observations}

The 8.46~GHz (3.5~cm) radio observations of most of the galaxies in the sample
were done with the VLA in two different runs, both part of the project AS\,713.
The first run was 7 hours long, during the CnB configuration, on 2001 June 24.
We observed galaxies with $\delta<-20^{\circ}$, or $\delta>50^{\circ}$
during this run. Most of the remaining galaxies were observed on the second
run, which was 13.5 hours long during the C configuration, on 2001 July 11. 
Besides these observations, we obtained data from the archive for those
sources which had already been observed with a similar configuration. We also
obtained archival 4.89~GHz (6~cm) data for the galaxies ESO\,350-G\,38,
NGC\,1741 and ESO\,400-G\,43, which did not have fluxes available in the
literature. New 1.49~GHz (20~cm) observations of NGC\,7552, not available on
the NVSS survey (Condon et al. 1998), were obtained on 2002 March 23,
during the A configuration, as part of the project AS\,721. We present in
Table~\ref{tabobs} the configurations in which the galaxies were observed,
the integration times and proposal codes from which they were obtained.

All the observations were done in continuum mode with 2 IF's of 50~MHz
bandwidth each, using the radio galaxies 3C\,48 and/or 3C\,286 as primary
calibrators. The phase calibration was done using calibrators from the NRAO
list, preferentially A-category calibrators closer than 10$^{\circ}$
from the galaxies. Most of the observations were done sandwiching $\sim$10
minutes observations of the galaxy with short $\sim$2-3~minutes
observations of a phase calibrator, repeating the process for 2 or 3 times.
The reductions followed standard AIPS techniques, which consisted of flagging
bad data points, setting the flux-density scale using the primary calibrators
and phase calibrating using the secondary calibrators. For those sources with
peak flux densities of $\sim$1~mJy or higher, we interactively self-calibrate
them two or three times in phase. The images were created using uniform
weighting. The final resolution of the 8.46~GHz images was of the order of
3\arcsec, and the maximum angular scales to which the observations were
sensitive is $\approx$3\arcmin. Given that some of the galaxies in our sample
have diameters of 3\arcmin\ or larger, like NGC\,6643, our observations may
have missed the short spacings, thus resolving out the more extended emission.
This is a limination of the observations, which can result in fluxes smaller
than the real ones, and can be solved by single dish 8.46~GHz observations.
Such problems were not an issue at 4.89~GHz and 1.49~GHz, since these
observations were done either with a single dish, or in the most compact
VLA configuration.

The noise of the images was determined in regions free from emission.
The total Stokes I fluxes [S(8.46~GHz)$_{int}$] were obtained by
integrating the regions brighter than 3$\sigma$ above the background level.
The fact that different galaxies can have a range of $\sim2$ in the 3$\sigma$
sensitivity level does not  have a significant impact on their integrated
fluxes, since most of the lower level flux does not cover large regions and
the integrated fluxes are dominated by the stronger regions. The errors in the
flux measurements were calculated by taking into account, in quadrature,
a 1.5\% uncertainty in the flux calibration and Poisson noise, which usually
dominates the errors.
For the galaxies for which we also have UV images, the radio images
were rotated, registered and their fluxes were measured inside a region
matching that observed in the UV [S(8.46~GHz)$_{match}$].
These values, and the beam sizes of the final images are presented in
Table~\ref{tabrad}. This Table also gives large beam 1.49~GHz and 4.89~GHz
fluxes obtained from the literature. Like in the case of the H$\alpha$
emission, we find that in most of the 8.46~GHz emission originates in the
region covered by the UV observations, with the median 
S(8.46~GHz)$_{match}$/S(8.46~GHz)$_{int}$ being 0.7.

\subsection{Optical and Infrared Data from the Literature}

Complementary to the images presented in this paper, we also collected
infrared, optical and ultraviolet data from the literature. The mid and
far infrared data, which will be used to determine the infrared luminosities
and star formation rates of the galaxies, were obtained from IRAS and ISO.
These observations are presented in Table~\ref{tabir}, they were obtained with
low spatial resolution, of the order of arminutes, which ensures that only
in the case of the most extended galaxies we may be missing a small amount of
emission. Nevertheless, even in cases like this the missing flux should be
negligible, since most of the emission is concentrated toward the nucleus.
The estimated uncertainty in the IRAS fluxes is of the order of 6\%, while
the errors from the ISO measurements are given in Table 5.

Ultraviolet, optical and near-infrared data are presented in Table~\ref{tabanc}.
These fluxes, combined with the mid and far-infrared ones, will be used to
calculate the bolometric luminosity of the galaxies. The ultraviolet fluxes
were obtained from IUE observations, which have an aperture of
10\arcsec $\times$20\arcsec\ (Kinney et al. 1993). The errors in these flux
measurements are given in Table~\ref{tabanc}. For cases where we have
both STIS 1457\AA\ and IUE 1482\AA\ observations, we use the ratio between
these two measurements to scale the IUE fluxes at longer wavelengths and
avoid aperture mismatch problems. For the galaxies that only have IUE
measurements, we do not try to apply any correction to the fluxes. This
can result in an uncertainty of a factor of 2 relative to the integrated
flux (see Section 5). We would also like to notice that for those galaxies
for which only IUE measurements are available we do not try to measured
matched aperture H$\alpha$ fluxes.

Optical (U, B, V and R band) and near infrared (J, H and K band) data were
obtained from broad band photometry, extrapolated to include the emission
from the entire galaxy (de Vaucouleurs et al. 1991; Jarrett et al. 2003).
The errors in the optical and near-infrared flux measurements are
10\% and 5\%, respectively.
These values are ideal for the comparison with far-IR data.
The near infrared fluxes are mostly from 2MASS observations, while the
optical ones were obtained from multiple sources in NED. These broad band
fluxes were converted to monochromatic fluxes using standard filter curves.

\section{Individual Objects}

\subsection{Galaxies with UV, H$\alpha$ and Radio Observations}

\subsubsection{MRK\,555}

Approximately 30\% of the UV emission originates in a circumnuclear ring
of $\sim$1\arcsec,
and we can also see other regions of emission, either diffuse or in the form of
clumps, at distances larger than 10\arcsec\ from the nucleus (Figure~\ref{fig1},
left column). The H$\alpha$ and radio images show strong emission associated
with these regions, besides emission along the spiral arms of the galaxy,
which was not covered by the UV image.

\subsubsection{NGC\,1569}

This is a nearby post-starburst galaxy (de Vaucouleurs, de Vaucouleurs, \& Pence
1974; Hodge 1974; Israel 1988) strongly affected by Galactic foreground
extinction (Figure~\ref{fig1}, right column). The UV image shows two large
clusters, separated by 8\arcsec\ along the NW-SE direction, as well as several
smaller clusters around them. The H$\alpha$ and radio emission trace each
other, however, they are anti correlated relative to the UV. Regions of strong
UV emission have weak H$\alpha$ and radio emission, indicating that the gas has
been evacuated by winds.

\subsubsection{NGC\,1667}

The nucleus of this galaxy hosts a low luminosity Seyfert 2 (Ho et al. 1997).
The images (left column of Figure~\ref{fig2}) show that the AGN contributes
very little to the overall UV, H$\alpha$ and Radio emission. Less than 1\% of
the UV emission originates from the nucleus of this galaxy (inner 2\arcsec),
where the emission may in part be due to nuclear radiation
scattered in our direction, or by circumnuclear star formation. Most of
the emission in this band comes from star forming regions along the
spiral arms of the galaxy. The H$\alpha$ image shows significantly
stronger emission at the nucleus, related to the Seyfert 2 nucleus of
this galaxy. It also shows emission from the star forming regions along
the spiral arms. The radio image shows strong emission related to the galaxy,
however, due to the fact that this emission is weak, we were not able to
self calibrate this image and some closure errors can still be seen.

\subsubsection{NGC\,1741}

The UV image shows two clumps of emission, associated with the two nuclei of
this galaxy, separated by $\sim$4\arcsec\ in the N-S direction (Figure~\ref{fig2},
right column). These two clumps are surrounded by smaller clusters of emission.
We can also see some diffuse emission extended to the NE of the northernmost
nucleus, as well as a tidal tail extending to the E. The H$\alpha$ emission
has a distribution similar to the UV, and also shows two tidal tails extending
for $\sim$1\arcmin\ to the SE and SW. The radio image shows emission
associated only with the double nucleus, where we can see that the stronger
radio source corresponds to the fainter UV nucleus. The UV image was registered
to the H$\alpha$ one, which is consistent with the radio.

\subsubsection{NGC\,4088}

The UV image is mostly empty, presenting only one strong clump of emission
and some diffuse emission (Figure~\ref{fig3}, left column). The H$\alpha$
image shows emission related to star forming regions
throughout the disk of the galaxy. The radio image shows emission related to
stronger star forming region of the galaxy, but is not deep enough to detect
the fainter, more diffuse regions seen in H$\alpha$. 

\subsubsection{NGC\,4214}

The UV image shows one strong clump of emission at the nucleus and another
clump on the eastern border of the image (Figure~\ref{fig3}, right column).
Like for NGC\,1569, the H$\alpha$ and radio emission trace each other, but
are very faint in regions of strong UV emission, indicating that the gas
have been evacuated from these regions. Notice that this galaxy is close
enough to allow us to resolve individual stars in the STIS image.

\subsubsection{MRK\,799}

The ultraviolet image of this galaxy (Figure~\ref{fig4} left column) shows some
faint emission, either diffuse or in the form of clumps, related to the spiral
arms of the galaxy. The H$\alpha$ and radio images trace this emission at the
nucleus and, besides the region of strong emission at 20\arcsec\ to the SE
and some faint emission to the N of the nucleus, outside the region covered by
the UV image.

\subsubsection{NGC\,5860}

The images of this starburst galaxy are presented in the right column of
Figure~\ref{fig4}. Notice that all the panels are shown on the same scale, so
we do not show the box indicating the region covered by the UV observations
in the H$\alpha$ image. The UV emission is distributed in two regions of similar
intensity, separated by 12\arcsec\ along the N-S direction. The H$\alpha$ image
shows strong emission associated with the southern component, but only fainter
emission related to the northern one, suggesting that it may be older than the
southern one. The radio image, however, only shows emission relate to the
southern component, in puzzling contradiction with the interpretation of the
H$\alpha$ image.

\subsubsection{ESO\,400-G\,43}

The left column of Figure~\ref{fig5} shows the images of this galaxy.
The UV image shows a very complicated structure, where we can identify several
individual clusters, as well as a detached clump of emission to the N.
The H$\alpha$ and radio images show structures similar to the UV, however, the
radio image does not show any emission related to the N component, probably due
to the sensitivity limit of the observations.

\subsubsection{NGC\,6217}

The UV image of this galaxy (Figure~\ref{fig5} right panel) shows strong
emission related to the nucleus and some fainter emission in the outer parts
of the image. The H$\alpha$ image detects strong nuclear emission, but also
detects a large number of fainter star forming regions along the spiral arms
of the galaxy, outside the region covered by the UV image. The radio image
is not sensitive enough to detect the fainter region of emission seen
in H$\alpha$, showing only a nuclear source and two faint point sources
in the field, one of which is not related to the galaxy. 

\subsubsection{NGC\,6643}

Figure~\ref{fig6}, left panel, presents the UV image of this galaxy, which
is mostly empty, showing only two faint point sources. The H$\alpha$ image
shows regions of star formation throughout the galaxy disk, while the radio one
shows only two point sources. Most of the diffuse emission seen in H$\alpha$
is not detected in the radio due to the sensitivity limit of the observations.

\subsubsection{MRK\,323}

The UV image of this galaxy shows only two point sources in the northern
border of the image (Figure~\ref{fig6} right column), which correspond to a
region of strong H$\alpha$ emission. The H$\alpha$ image also shows some
fainter emission around the nucleus of the galaxy, which is related to the
strongest region of radio emission.

\subsubsection{MRK\,332}

The UV image of this spiral galaxy (Figure~\ref{fig7}) presents strong
emission associated to the nucleus, as well as several star forming regions
along the spiral arms. The H$\alpha$ emission traces what is seen in the UV,
and also shows strong emission along the spiral arms, outside the region
covered by the UV image. The radio shows only a strong nuclear point
source, and some faint emission corresponding to the spiral arms.

\subsection{Galaxies with UV and Radio Observations}

\subsubsection{IC\,1623}

This is an interacting galaxy with a double nucleus (Knop et al. 1994).
The UV image (Figure~\ref{fig8} top) shows only the western component, while
the eastern one is completely enshrouded by dust, being visible only at
longer wavelengths. The radio image shows emission related to both components,
however, the stronger source is related to the eastern, more obscured,
structure. A detailed study of the UV image of this galaxy is presented
by Goldader et al. (2002), found that both components have similar bolometric
luminosities.

\subsubsection{NGC\,3079}

This is an edge on spiral galaxy, with several regions of star formation along
the disk (Cecil et al. 2001). The UV image (Figure~\ref{fig8} middle panel)
is mostly empty, showing only traces of emission, due to the high amount of
extinction along the line of sight. The radio image shows a strong point source
at the nucleus and a lobe at 20\arcsec\ NE, both related to a hidden AGN
(Duric et al. 1983). The radio image also shows strong emission along
P.A.$=-15^{\circ}$, related to star forming regions along the galaxy disk.
The H$\alpha$ image from Cecil et al. (2001) shows strong emission along
the disk, as well as a nuclear outflow, related to the radio lobe.

\subsubsection{NGC\,4861}

We present in the bottom panel of Figure~\ref{fig8} the UV image of this
nearby starburst galaxy. The emission is divided into two large clumps,
one corresponding to the nucleus, in the center of the image, which is
composed of several smaller clusters. The second clump is more diffuse,
located 9\arcsec\ to the NE. The radio image shows strong emission related
to the nucleus, but nothing towards the NE structure.
Dottori et al. (1994) presented an H$\alpha$ image of this galaxy,
which shows emission related with all the structures seen in the UV.

\subsubsection{NGC\,5383}

The UV image of this galaxy (Figure~\ref{fig9} top left panel) shows several
clusters, as well as some diffuse emission, extended for $\sim$20\arcsec\ in
the E-W direction in the form of a ring, or inner spiral arms. Comparing this
image with the H$\alpha$ one presented by Sheth et al. (2000), we see that
they trace each other, with the regions of strong H$\alpha$ emission following
the ones with strong UV. The overall structure of the radio emission (right
panel) is similar to the H$\alpha$ and UV. However, when they are compared
in detail we find that the peak of the radio emission is located in a region
with little, or no UV and H$\alpha$ emission. This enhanced radio emission
could be related to a region of star formation hidden by dust, or to the
nucleus of the galaxy. The latter possibility implies that this galaxy harbors
a low luminosity AGN. If this is true, this AGN should be completely hidden,
since Ho, Filippenko \& Sargent (1997) classified this galaxy as HII.

\subsubsection{NGC\,5676}

Figure~\ref{fig9}, middle panel, presents the UV and radio images of this
galaxy. Only faint emission is detected in the UV. The radio image also shows
faint emission in the same region covered by the UV observations, indicating
that the faint emission is due to low star formation rate rather than
obscuration. A region with stronger radio emission is seen to the N of the
nucleus, and outside of the STIS field of view.

\subsubsection{NGC\,5713}

The UV and radio images of this galaxy are presented in the bottom panel of
Figure~\ref{fig9}. The UV emission shows several associations, with the
strongest ones located to the northern part of the image. The radio emission
is concentrated in a string of knots, aligned with the structures seen
in the northern part of the UV image. The two easternmost knots correspond
to regions of emission in the UV, but the third one, around right ascension
14:40:10.7, is related to a region without UV emission. Conversely, analyzing
the regions to the west of the UV image, we can detect strong UV emission,
but only weak radio emission is associated with these structures.

\subsubsection{UGC\,11284}

This is an interacting galaxy where the eastern component presents stronger
H$\alpha$ emission than the western one (Bushouse 1986). The UV observations
were centered on the eastern component, but we detect only a faint compact
source coincident with the radio peak and some diffuse emission to the east,
which is also visible in the radio
(Figure~\ref{fig10}). The radio image shows the two galaxies and some
faint emission between them. Like in the case of H$\alpha$, the eastern
component is the strongest one, suggesting that the star forming region is
highly reddened, or the possible presence of a hidden AGN. The integrated
8.46~GHz flux given in Table~\ref{tabrad} corresponds to the flux of the two
galaxies, to be consistent with the region covered by the far-IR observations.

\subsection{Galaxies with H$\alpha$ and Radio Observations}

\subsubsection{ESO\,350-G\,38}

The H$\alpha$ and radio emission of this starburst galaxy (Figure~\ref{fig11}
top) show a good agreement. Both images show a Y shaped structure at the
nucleus, with the strongest emission coming from the center. The H$\alpha$
image shows some diffuse emission surrounding this Y shaped structure, which
is not seen in the radio. The UV image of this galaxy (Kunth et al. 2003) shows
strong emission related to the E and S component, but only faint emission
related to the strong nuclear source seen in H$\alpha$ and radio.

\subsubsection{NGC\,232}

The middle panels of Figure~\ref{fig11} present the H$\alpha$ and radio
images of this galaxy, which shows strong H$\alpha$ emission concentrated
at the nucleus and in a blob
at 3.5\arcsec\ W of it. We also see more diffuse emission extended along
P.A.$\sim-40^{\circ}$, which is in the direction at which the radio emission
is extended.

\subsubsection{NGC\,337}

This galaxy presents diffuse H$\alpha$ emission distributed throughout the
disk (Figure~\ref{fig11} bottom panel), as well as some clumps with stronger
emission. The structure of the radio emission is similar to that seen in
H$\alpha$, however, a significant part of the diffuse emission is resolved
out in the radio image, because the observations do not have short spacings.

\subsubsection{NGC\,5161}

The H$\alpha$ image of this galaxy presents small star forming regions along
the spiral arms, as well as some faint emission related to the nucleus
(Figure~\ref{fig12} top). The radio image shows emission at the nucleus, and
some faint emission related to the host galaxy.

\subsubsection{NGC\,6090}

Figure~\ref{fig12}, middle panel, presents the H$\alpha$ and radio images of
this galaxy, which have similar structures. The emission is divided into
two major knots, separated by $\sim$6\arcsec\ along the NE-SW direction. The
NE structure can be divided into two components separated by $\sim$2\arcsec\
in the N-S direction.

\subsubsection{NGC\,7496}

This galaxy has a low luminosity AGN in the nucleus, classified as a Seyfert 2
(Kinney et al. 1993),
nevertheless it also shows strong circumnuclear star formation. The bottom
panels of Figure~\ref{fig12} present the H$\alpha$, which shows several blobs
of emission around the nucleus, a strong region of emission at 40\arcsec\ NW,
as well as some faint star forming regions along the disk. The radio image
shows emission related to the nuclear and circumnuclear region and a faint
point source at $\sim$1\arcmin\ N.

\subsubsection{NGC\,7552}

This is a nuclear starburst galaxy. The H$\alpha$ and radio images
(Figure~\ref{fig13} top) show the presence of a circumnuclear ring, as
well as star formation along the bar of the galaxy (similar structures
were detected by Hameed \& Devereux 1999). The H$\alpha$ image
also shows some emission along the spiral arms, which is too faint to be
detected by our radio observations. 

\subsubsection{TOL\,1924-416}

The bottom panel of Figure~\ref{fig13} shows the H$\alpha$ and radio images
of this galaxy. The emission in these bands is very similar, two blobs
separated by 6\arcsec\ along the E-W direction. The H$\alpha$ image shows some
diffuse emission surrounding these structures, which is not seen in the radio.
A detailed high spatial resolution study of this galaxy was presented
by \"Ostlin, Bergvall \& R\"onnback (1998), who detected a large number of
star clusters in this galaxy. These starbursts show peaks in the age
distribution, with the younger ones being found in the starburst region.
The galaxy is in a merging state, which probably caused the starburst.
At 10\arcsec\ SW from the westernmost H$\alpha$ blob we see some image
defects due to residuals from the continuum subtraction of a foreground star.

\subsection{Galaxies with Observations in a Single Band}

\subsubsection{IC\,1586}
This galaxy was observed only in the radio. The 3.6~cm image
(Figure~\ref{fig14} top left), shows two faint blobs, separated
by $\sim$7\arcsec\ in the NE-SW direction.

\subsubsection{NGC\,1155}
The radio emission of this galaxy presents only a compact
source (Figure~\ref{fig14} top right)

\subsubsection{UGC\,2982}
The radio emission of this galaxy is presented in Figure~\ref{fig14}
(middle left panel). It is diffuse and extended for approximately 45\arcsec\
along the E-W direction.

\subsubsection{NGC\,1614}
The 3.6~cm image of this galaxy consists of a strong point source,
surrounded by faint emission, extended by 25\arcsec\ in the E-W direction
(Figure~\ref{fig14} middle right). The structure of the radio emission is
similar to that seen in H$\alpha$ (Armus, Heckman \& Miley 1990).

\subsubsection{NGC\,3690}
Figure~\ref{fig14} (bottom left panel) presents the radio image of this
galaxy. The bulk of the emission is concentrated in a region with a diameter
of 50\arcsec. The emission in this region consists of several blobs and some
faint emission surrounding it. Fainter emission can be seen towards
the NW. The radio and H$\alpha$ emission of this galaxy have similar
structures (Armus et al. 1990).

\subsubsection{NGC\,4100}
Most of the radio emission of this galaxy (Figure~\ref{fig14} bottom right) is
very faint and diffuse, close to the detection limit of the observation. A region
with stronger emission is seen around the nucleus.

\subsubsection{NGC\,5054}
The radio image of this galaxy, presented in the top panel of
Figure~\ref{fig15}, shows stronger emission at the nucleus, surround by
a region of fainter and diffuse emission, with a radius of 40\arcsec.

\subsubsection{NGC\,7673}
Figure~\ref{fig15} (middle panel) shows the radio emission of this galaxy,
which has the form of a ring, composed of 3 knots, more or less centered
around the nucleus. Homeier \& Gallagher (1999) presented the H$\alpha$
image of this galaxy, which has a structure similar to the one seen in the
radio. Their image also shows a region of strong emission between the two
brighter radio structures, which does not seem to have a counterpart
in the radio.

\subsubsection{NGC\,7714}
The radio emission of this galaxy presents a strong point source at the
nucleus (Figure~\ref{fig15} bottom panel) and some diffuse emission around
it, extending by approximately 30\arcsec\ in the NW-SE direction. This
structure is similar to
the one seen in H$\alpha$ by Gonz\'alez Delgado et al. (1995).

\subsubsection{NGC\,1672}
The UV image of this galaxy is presented in the left panel of
Figure~\ref{fig16}. It shows a structure resembling a half ring,
with several clusters of stars along it. The regions of stronger UV
emission are coincident with those of strong H$\alpha$ emission
seen by Storchi-Bergmann, Wilson \& Baldwin (1996).

\subsubsection{NGC\,5669}
The UV image of this galaxy (Figure~\ref{fig16} right panel) shows enhanced
emission along the bar (NE-SW), as well as some emission along the inner
spiral arms, which are seen in the form of a ring, close to the
borders of the image. We obtained VLA 8.46~GHz data for this galaxy, but
could not detected any radio emission. Lower resolution ($\sim$45\arcsec\
beam), lower frequency (1.4~GHz) observations from the NVSS catalogue
(Condon et al. 1998), detected this source. Nevertheless, higher resolution
observations ($\sim$4.5\arcsec\ beam) at the same frequency, from the FIRST
catalogue (Becker, White \& Helfand 1995), did not detect it either. This
suggests that the star formation in this galaxy is faint and widespread,
resulting in radio emission below the detection threshold of our 8.46~GHz
observations.

\subsubsection{NGC\,6753}
The left panel of Figure~\ref{fig17} presents the H$\alpha$ image of this
galaxy, which is distributed throughout the disk and spiral arms. The
latter can be seen as a ring structure extended along the NE-SW direction.
Notice that the two bright sources to the south of the galaxy are residuals
due to foreground stars, left over from the subtraction of the continuum.

\subsubsection{NGC\,6810}
The H$\alpha$ image of this galaxy (Figure~\ref{fig17} right panel) shows
strong emission, extended for $\sim$1\arcmin\ along the N-S direction, in a
rectangular region. This emission is composed of several individual knots.
Diffuse emission, extending over a region of 80\arcsec\ in diameter, is seen
along the SE-NW direction, possibly related to outflows. Another interesting
structure that can be seen in this image is a half ring of small knots,
to the E of the rectangular region described above. This ring extends along
the major axis of the galaxy, has a diameter of $\sim$100\arcsec\ in the N-S
direction, and lies along the border of a dust lane. Observations by
Hameed \& Devereux (1999) found similar structures.

\section{Discussion and Summary}

In this paper we present the compilation of a database of multiwavelength
data for a sample of 41 nearby star forming galaxies. We present the
observations and reductions of UV images for 22 of these galaxies,
H$\alpha$ images for 23 galaxies, and radio 8.46~GHz images for 39 galaxies. 
We also obtained H$\alpha$ data from the literature for another 14 galaxies
which we could not observe, as well as complementary multifrequency radio,
infrared and optical data for all the galaxies in the sample.

A preliminary comparison between the UV, H$\alpha$ and radio images presented
in this paper shows that, overall, on a larger scale, there is a good agreement
in the spatial distribution of the different star formation indicators. Regions
of intense star formation usually show up as strong regions of emission at all
wavelengths. The relation between the different star formation indicators will
be explored in more detail in an accompanying paper, but the current results
already suggest that these 3 indicators trace each other, at least within the
physical parameter space covered by our sample. However, an important
effect that should be taken into account when dealing with shorter wavelength
measurements (UV and H$\alpha$)  is dust obscuration, which can easily absorb,
or even hide entire regions of a galaxy. The most noticeable example of such
effect is IC\,1623, where the eastern
component can be detected only in the radio, while the western one can be seen
at both UV and radio. Similar results can also be seen in other galaxies.

Despite the good agreement between the different wavebands described above,
a detailed inspection of individual sources shows that in some cases there
are deviations from this pattern, which cannot be attributed to dust. One such
example is NGC\,6643, where the UV image is almost empty, while the H$\alpha$
one shows several regions of emission throughout the disk of the galaxy,
and the radio shows only two small regions of emission. The differences
from the expected picture can be attributed in part to the fact that the star
formation in this galaxy is mostly faint and diffuse, so the UV emission
could have easily been absorbed by dust. In the case of the radio 8.46~GHz
emission, the observations were not sensitive enough to detect the faint
diffuse emission along the disk of the galaxy, but this emission was clearly
detected by the large beam observations at lower frequencies. Two other
galaxies which present an interesting behavior are NGC\,1569 and NGC\,4214.
Based on a visual inspection of their UV images we can clearly see bright
star clusters, which indicates that they are not strongly absorbed.
Nevertheless, when these images are compared to the H$\alpha$ and radio
ones, we find that the emission in these wavebands is displaced relative to
the clusters, suggesting that the region around them was evacuated by winds.
Although not all galaxies show these effects, they can have a significant
impact on the determination of star formation rates.

The broad wavelength range spanned by our data is better appreciated in the
left panel of Figure~\ref{fig18}, where the UV--to--radio SEDs of 29 of our
sample galaxies are plotted with normalization of unity at 1.49~GHz (20~cm).
In the case of galaxies for which 1.49~GHz was not available we extrapolated
the flux from higher frequencies, assuming S$_{\nu}\propto\nu^{-1}$. The 29
galaxies are those for which at least one ISO far--infrared data point
beyond 100~$\mu$m is available. In the case of NGC\,4088 we exclude the
measurement at 170$\mu$m, which has calibration problems, and use only
the one at 180$\mu$m (Table~\ref{tabir}). All data are integrated fluxes,
except for the UV points, which were measured with a smaller field of
view. We did not try to apply corrections to the UV data for aperture mismatch.
In most cases, the mismatch between the UV values and the integrated values
at longer wavelengths is at the level of a factor of 2 or less. This factor
is inferred from the comparison of the H$\alpha$ and 8.46~GHz fluxes measured
inside an aperture matching the UV one, and their corresponding integrated
values (Sections 3.2 and 3.3). Thus, the SEDs plotted in Figure~\ref{fig18}
can be considered representative of each galaxy's SED even in the UV.

This collection of normalized SEDs of normal star--forming galaxies readily
shows a number of characteristics. (1) The range in far--infrared luminosity
is fairly modest, once the SEDs are normalized to the radio,a factor of 5--6,
clearly a reflection of the tight FIR--radio correlation found for galaxy
populations (de Jong et al. 1985, Helou et al. 1985). (2) The range
in FIR luminosity is much smaller than the factor $\sim$200 spanned by
the UV and $\sim$100 spanned by the optical; the large UV and optical
variation is likely a combination of variations in the stellar
population content and the dust obscuration, with the latter
contributing to the larger spread observed in the UV. (3) Relatively
`blue' SEDs at optical wavelengths (those for which
$\nu~F(\nu)_{optical}\approx\nu~F(\nu)_{FIR}$) do not necessarily
correspond to blue galaxies in the UV as well; this likely reflects
the stellar population content of the galaxy, where the
`optically--bright' populations are likely evolved stars, no longer
UV--bright and contributing to the star formation. 

The thicker dashed lines in the left panel of Figure~\ref{fig18} identify the
two basic templates that can be recognized in our modest collection of SEDs
(the average values are given in Table~\ref{tabsed}): (a) the FIR-dominated
SED (REDSED), created averaging 11 galaxies, where the observed UV--optical
emission is heavily affected by dust, and provides only a modest contribution
to the galaxy energy budget; (b) the optical$\sim$FIR SED (BLUESED), created
averaging 18 galaxies, where the evolved stellar populations are as important
as the star-forming (FIR-emitting) population for the galaxy's energy budget.
Due to their large wavelength range, these two basic templates can be used to
gain insights in the properties of intermediate and large redshift galaxies. 

We also present, in the right panel of Figure~\ref{fig18}, the comparison
between the SEDs created in this paper with those of high and low reddening
starbursts (SBH and SBL, respectively) from Schmitt et al. (1997).
We can see that there is a good agreement between the two red SEDs, SBH and
REDSED, over the entire frequency range. However, in the case of the blue
ones, SBL and BLUESED, the agreement is good only in the radio--FIR
range, with the BLUESED presenting higher fluxes in the optical-near
infrared range, but smaller flux in the ultraviolet. These differences can
be attributed to aperture effects. The SEDs created with the current database
include the entire galaxy, with the exception of the ultraviolet flux.
In the case of the Schmitt et al. (1997) SEDs the ultraviolet, optical and
near-infrared measurements were obtained with apertures similar to that
of IUE ($\sim10^{\prime\prime}\times20^{\prime\prime}$), while the radio and
infrared measurements included the entire galaxy. This aperture difference
is not a significant problem for the infrared dominated SEDs, since they
usually are dominated by a reddened nuclear starburst. Nevertheless, in
the case of larger galaxies, without a nuclear starburst, older stars 
make a significant contribution to the integrated light of the galaxy, 
resulting in the similar optical--to--FIR peak ratio observed in the BLUESED,
and in larger optical--to--UV ratios than observed in the low reddening
starburst SED of Schmitt et al. (1997). 

\acknowledgements

This work was partially supported by the NASA grants HST-GO-8721 and
NAG5-8426.
The National Radio Astronomy Observatory is a facility of the
National Science Foundation, operated under cooperative agreement by Associated
Universities, Inc. This research made use of the NASA/IPAC Extragalactic
Database (NED), which is operated by the Jet Propulsion Laboratory, Caltech,
under contract with NASA. We also used the Digitized Sky Survey, which was
produced at the Space telescope Science Institute under U.S. Government
grant NAGW-2166. HRS would like to acknowledge the NRAO Jansky Fellowship
program for support during most of the stages of this project. HRS would also
like to thank the Spitzer Science Center,
and the Space Telescope Science Institute visitor programs for their support.
UV observations were obtained with the NASA/ESA Hubble Space Telescope at the
Space Telescope Science Institute, which is operated by the Association of
Universities for Research in Astronomy, Inc., under NASA contract NAS5-26555.
Basic research at the US Naval Research Laboratory is supported by the Office
of Naval Research. We would like to thank the referee for comments that helped
us improve this paper.

\clearpage

\begin{figure}
\epsscale{0.8}
\caption{The radio 8.46~GHz (top panels), H$\alpha$ (middle panels) and
ultraviolet (bottom panels) images of MRK\,555 (left column) and NGC\,1569
(right column). The H$\alpha$ and radio images are on the same scale, while
the region covered by the ultraviolet image is shown as a box on top of the
H$\alpha$ contours. The contour levels of the radio images presented in this
paper start at the 3$\sigma$ level above the background and increase in powers
of 2 (2$^n\times3\sigma$), while for the H$\alpha$ images they start at the
3$\sigma$ level and increase in powers of 3 (3$^n\times3\sigma$).}
\label{fig1}
\end{figure}

\begin{figure}
\epsscale{1}
\caption{Same as Figure 1 for NGC\,1667, left, and NGC\,1741, right.}
\label{fig2}
\end{figure}

\begin{figure}
\caption{Same as Figure 1 for NGC\,4088, left, and NGC\,4214, right.}
\label{fig3}
\end{figure}

\begin{figure}
\caption{Same as Figure 1 for MRK\,799, left, and NGC\,5860, right.
All NGC\,5860 images are presented on the same scale, so we do not show
the region covered by the UV emission in the H$\alpha$ image.}
\label{fig4}
\end{figure}

\begin{figure}
\caption{Same as Figure 1 for ESO\,400-G\,43, and NGC\,6217, right.}
\label{fig5}
\end{figure}

\begin{figure}
\caption{Same as Figure 1 for NGC\,6643, left and MRK\,323, right.}
\label{fig6}
\end{figure}

\begin{figure}
\caption{Same as Figure 1 for MRK\,332.}
\label{fig7}
\end{figure}

\begin{figure}
\epsscale{0.8}
\caption{The left column shows the ultraviolet, while the right one shows the
radio 8.46~GHz images of IC\,1623 (top row), NGC\,3079 (middle row) and NGC\,4861
(bottom row). In the case of IC\,1623, both images are shown on the same scale,
while in the case of NGC\,3079 and NGC\,4861 the region covered by the ultraviolet
image is shows as a box superimposed on the contours.}
\label{fig8}
\end{figure}

\begin{figure}
\caption{Same as Figure 8 for NGC\,5383 (top row), NGC\,5676 (middle row)
and NGC\,5713 (bottom row). The region covered by the ultraviolet image
is shown as a box superimposed on the contours.}
\label{fig9}
\end{figure}

\begin{figure}
\epsscale{1}
\caption{Same as Figure 8 for UGC\,11284.}
\label{fig10}
\end{figure}

\begin{figure}
\caption{The left column shows the H$\alpha$, while the right one shows the
radio 8.46~GHz images of ESO\,350-G\,38 (top row), NGC\,232 (middle row),
and NGC\,337 (bottom row). The H$\alpha$ and radio image are on the same
scale.}
\label{fig11}
\end{figure}

\begin{figure}
\caption{Same as Figure 11 for NGC\,5161 (top row), NGC\,6090 (middle row),
and NGC\,7496 (bottom row).}
\label{fig12}
\end{figure}

\begin{figure}
\caption{Same as Figure 11 for NGC\,7552 (top) and TOL\,1924-416 (bottom).}
\label{fig13}
\end{figure}

\begin{figure}
\caption{Radio 8.46~GHz images of IC\,1586 (top left), NGC\,1155 (top right),
UGC\,2982 (middle left), NGC\,1614 (middle right), NGC\,3690 (bottom left)
and NGC\,4100 (bottom right).}
\label{fig14}
\end{figure}

\begin{figure}
\caption{Same as Figure 14 for NGC\,5054 (top), NGC\,7673 (middle) and
NGC\,7714 (bottom).}
\label{fig15}
\end{figure}

\begin{figure}
\caption{Ultraviolet images of NGC\,1672 (left) and NGC\,5669 (right).}
\label{fig16}
\end{figure}

\begin{figure}
\caption{H$\alpha$ images of NGC\,6753 (left) and NGC\,6810 (right).}
\label{fig17}
\end{figure}

\begin{figure}
\plotone{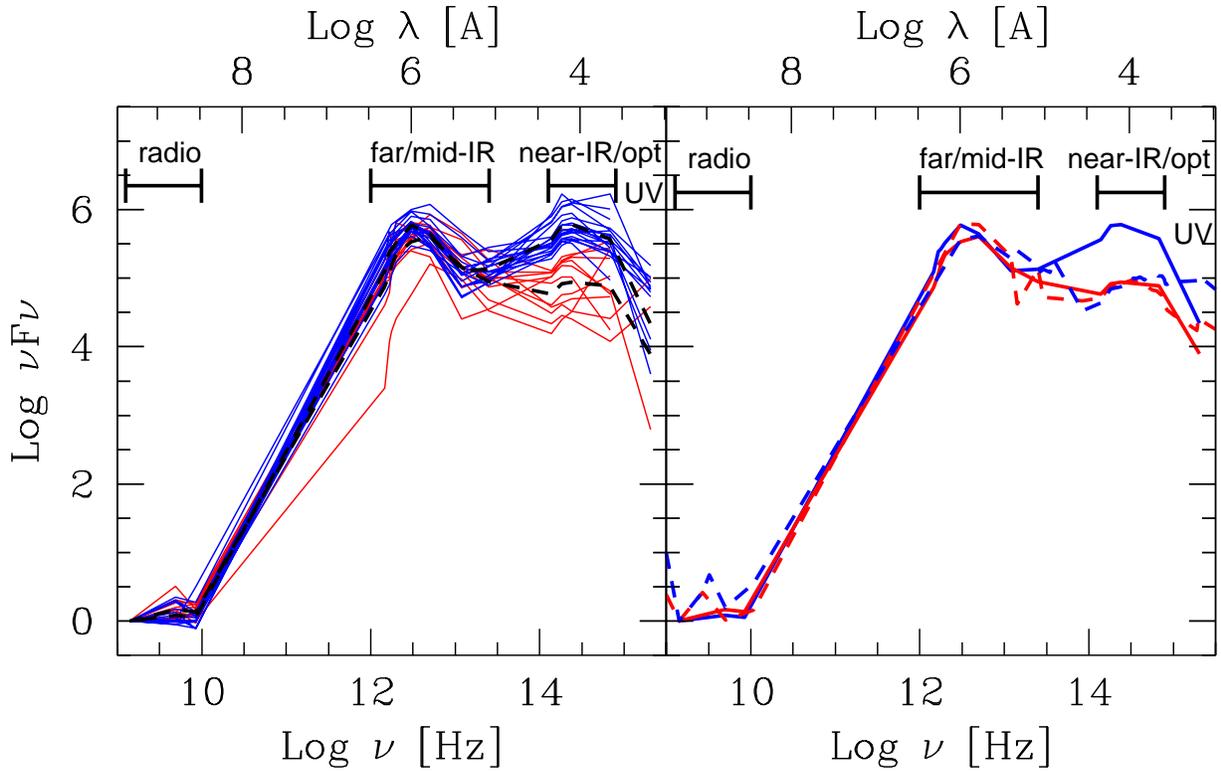}
\caption{Left: spectral energy distributions of the 29 galaxies for which
we have at least one ISO band. Galaxies with SEDs dominated by the far infrared
emission are presented in red (REDSED), while those with comparable optical
and far-infrared fluxes are presented in blue (BLUESED). The thick dashed
lines indicate the average SED of each group (Table~\ref{tabsed}). Right:
BLUESED and REDSED, solid blue and red lines respectively, are compared to
the low (SBL) and high (SBH) reddening SEDs from Schmitt et al. (1997), blue
and red dashed lines respectively. We show in both panels the wavelength scale
in the top axis and also indicate the different waveband regions inside the
panels.}
\label{fig18}
\end{figure}

\clearpage
\begin{deluxetable}{lrrrrrrl}
\tabletypesize{\scriptsize}
\tablewidth{0pc}
\tablecaption{Sample Characteristics}
\tablehead{\colhead{Name}&
\colhead{$\alpha$(J2000.0)}&
\colhead{$\delta$(J2000.0)}&
\colhead{Vel.}&
\colhead{Dist.}&
\colhead{a$\times$b}&
\colhead{E(B-V)}&
\colhead{Morph.}\\
\colhead{}&
\colhead{}&
\colhead{}&
\colhead{(km s$^{-1}$)}&
\colhead{(Mpc)}&
\colhead{(\arcmin)}&
\colhead{}&
\colhead{}\\
\colhead{(1)}&
\colhead{(2)}&
\colhead{(3)}&
\colhead{(4)}&
\colhead{(5)}&
\colhead{(6)}&
\colhead{(7)}&
\colhead{(8)}}
\startdata
ESO\,350-G\,38&  0 36 52.5& -33 33 19 & 6132 & 81.8 &0.5$\times$0.3 & 0.01 & Merger  \\
NGC\,232      &  0 42 45.8& -23 33 41 & 6682 & 89.1 &1.0$\times$0.8 & 0.01 & SB(r)a pec  \\
MRK\,555      &  0 46 05.6& -01 43 25 & 4156 & 55.4 &1.4$\times$1.2 & 0.02 & SA(rs)b pec  \\
IC\,1586      &  0 47 56.1&  22 22 21 & 5963 & 79.5 &0.3$\times$0.3 & 0.02 & BCG  \\
NGC\,337      &  0 59 50.3& -07 34 44 & 1702 & 20.7 &2.9$\times$1.8 & 0.08 & SB(s)d  \\
IC\,1623      &  1 07 47.2& -17 30 25 & 6016 & 80.2 &0.7$\times$0.4 & 0.01 & Merger  \\
NGC\,1155     &  2 58 13.0& -10 21 04 & 4549 & 60.7 &0.8$\times$0.7 & 0.04 & Compact  \\
UGC\,2982     &  4 12 22.6&  05 32 51 & 5273 & 70.3 &0.9$\times$0.4 & 0.14 & Sm  \\
NGC\,1569     &  4 30 49.3&  64 50 54 &   40 &  1.6 &3.6$\times$1.8 & 0.51 & IBm  \\
NGC\,1614     &  4 34 00.0& -08 34 45 & 4681 & 62.4 &1.3$\times$1.1 & 0.06 & SB(s)c pec  \\
NGC\,1667     &  4 48 37.1& -06 19 13 & 4459 & 59.5 &1.8$\times$1.4 & 0.06 & SAB(r)c  \\
NGC\,1672     &  4 45 42.1& -59 14 57 & 1155 & 14.5 &6.6$\times$5.5 & 0.00 & SB(r)bc  \\
NGC\,1741     &  5 01 37.9& -04 15 35 & 3956 & 52.8 &1.5$\times$1.0 & 0.06 & Merger  \\
NGC\,3079     & 10 01 57.8&  55 40 49 & 1182 & 20.4 &7.9$\times$1.4 & 0.00 & SB(s)c  \\
NGC\,3690     & 11 28 32.6&  58 33 47 & 3121 & 41.6 &3.0$\times$3.0 & 0.00 & Merger  \\
NGC\,4088     & 12 05 35.6&  50 32 32 &  825 & 17.0 &5.8$\times$2.2 & 0.00 & SAB(rs)bc  \\
NGC\,4100     & 12 06 08.7&  49 34 56 & 1138 & 17.0 &5.4$\times$1.8 & 0.01 & (R')SA(rs)bc  \\
NGC\,4214     & 12 15 40.0&  36 19 27 &  313 &  3.5 &8.5$\times$6.6 & 0.00 & IAB(s)m  \\
NGC\,4861     & 12 59 00.3&  34 50 48 &  880 & 17.8 &4.0$\times$1.5 & 0.01 & SB(s)m:  \\
NGC\,5054     & 13 16 58.4& -16 38 03 & 1630 & 27.3 &5.1$\times$3.0 & 0.03 & SA(s)bc  \\
NGC\,5161     & 13 29 14.4& -33 10 29 & 2247 & 33.5 &5.6$\times$2.2 & 0.04 & SA(s)c:  \\
NGC\,5383     & 13 57 05.0&  41 50 44 & 2333 & 37.8 &3.2$\times$2.7 & 0.00 & (R')SB(rs)b:pec  \\
MRK\,799      & 14 00 45.8&  59 19 44 & 3157 & 42.1 &2.2$\times$1.1 & 0.00 & SB(s)b  \\
NGC\,5669     & 14 32 44.1&  09 53 24 & 1387 & 24.9 &4.0$\times$2.8 & 0.01 & SAB(rs)cd  \\
NGC\,5676     & 14 32 46.8&  49 27 30 & 2237 & 34.5 &4.0$\times$1.9 & 0.01 & SA(rs)bc  \\
NGC\,5713     & 14 40 11.3& -00 17 26 & 1872 & 30.4 &2.8$\times$2.5 & 0.03 & SAB(rs)bc pec  \\
NGC\,5860     & 15 06 33.4&  42 38 29 & 5520 & 73.6 &1.0$\times$1.0 & 0.01 & Peculiar  \\
NGC\,6090     & 16 11 40.8&  52 27 27 & 8953 & 119.4&1.0$\times$1.0 & 0.00 & Peculiar  \\
NGC\,6217     & 16 32 39.3&  78 11 53 & 1544 & 23.9 &3.0$\times$2.5 & 0.04 & (R)SB(rs)bc  \\
NGC\,6643     & 18 19 46.0&  74 34 09 & 1686 & 25.5 &3.8$\times$1.9 & 0.06 & SA(rs)c  \\
UGC\,11284    & 18 33 35.5&  59 53 20 & 8650 & 115.3&2.0$\times$2.0 & 0.05 & Merger  \\
NGC\,6753     & 19 11 23.4& -57 02 56 & 3073 & 41.0 &2.5$\times$2.1 & 0.06 & (R')SA(r)b  \\
TOL\,1924-416 & 19 27 58.0& -41 34 28 & 2863 & 38.2 &0.8$\times$0.4 & 0.08 & Peculiar  \\
NGC\,6810     & 19 43 34.2& -58 39 21 & 1888 & 25.3 &3.2$\times$0.9 & 0.04 & SA(s)ab:sp  \\
ESO\,400-G\,43& 20 37 41.8& -35 29 04 & 6032 & 80.4 &0.5$\times$0.4 & 0.02 & Compact  \\
NGC\,7496     & 23 09 47.0& -43 25 40 & 1623 & 20.1 &3.3$\times$3.0 & 0.01 & (R')SB(rs)bc  \\
NGC\,7552     & 23 16 11.0& -42 34 59 & 1568 & 19.5 &3.4$\times$2.7 & 0.01 & (R')SB(s)ab  \\
MRK\,323      & 23 20 22.7&  27 18 56 & 4404 & 58.7 &1.0$\times$0.7 & 0.05 & SBc  \\
NGC\,7673     & 23 27 41.3&  23 35 23 & 3581 & 47.8 &1.3$\times$1.2 & 0.04 & (R')SAc pec.  \\
NGC\,7714     & 23 36 14.1&  02 09 18 & 2925 & 39.0 &1.9$\times$1.4 & 0.04 & SB(s)b:pec  \\
MRK\,332      & 23 59 25.6&  20 45 00 & 2568 & 34.2 &1.4$\times$1.3 & 0.04 & SBc  \\
\enddata
\tablenotetext{a}{Column 1: galaxy name;
columns 2 and 3: right Ascension and Declination, respectively, in J2000.0
coordinates;
column 4: radial velocities relative to the Local Group;
column 5: distances, which were obtained from Tully (1988) for
the nearest galaxies, or calculated from the radial velocities
assuming H$_0=75$ km s$^{-1}$ Mpc$^{-1}$;
column 6: the galaxies' major and minor axis diameters (D$_{25}$);
column 7: galactic foreground reddening, obtained preferentially from
Burstein \& Heiles (1982), or from Schlegel, Finkbeiner \& Davies (1998)
when the previous one was not available;
column 8: morphological types.}
\label{tabcar}
\end{deluxetable}

\begin{deluxetable}{lrrrrrlrrrr}
\tabletypesize{\scriptsize}
\tablewidth{0pc}
\tablecaption{Observations}
\tablehead{
\colhead{}&
\multicolumn{2}{c}{Ultraviolet}&
\colhead{}&
\multicolumn{3}{c}{H$\alpha$}&
\colhead{}&
\multicolumn{3}{c}{Radio}\\
\cline{2-3} \cline{5-7} \cline{9-11}\\
\colhead{Name}&
\colhead{Dataset}&
\colhead{Exp.}&
\colhead{}&
\colhead{Observatory}&
\colhead{Exp.}&
\colhead{Filter}&
\colhead{}&
\colhead{Config.}&
\colhead{Exp.}&
\colhead{Prop.}\\
\colhead{(1)}&
\colhead{(2)}&
\colhead{(3)}&
\colhead{}&
\colhead{(4)}&
\colhead{(5)}&
\colhead{(6)}&
\colhead{}&
\colhead{(7)}&
\colhead{(8)}&
\colhead{(9)}}
\startdata
ESO\,350-G\,38&\nodata       &\nodata & &CTIO    &3600    &6680/100 & &BnA     &5840    &AJ\,176 \\
              &\nodata       &\nodata & &\nodata &\nodata &\nodata  & &BnA     &3750$^d$&AJ\,176 \\
NGC\,232      &\nodata       &\nodata & &CTIO    &3600    &6680/100 & &CnB     &1670    &AS\,713 \\
MRK\,555      &O63X04WWQ     &1320    & &WIYN-2  &720     &KP\,1494 & &C       &1700    &AS\,713 \\
IC\,1586      &\nodata       &\nodata & &\nodata &\nodata &\nodata  & &C       &1650    &AS\,713 \\
NGC\,337      &\nodata       &\nodata & &CTIO    &3600    &6600/75  & &C       &1680    &AS\,713 \\
IC\,1623      &O5CU02010$^a$ &3051    & &\nodata &\nodata &\nodata  & &CnB     &1680    &AS\,713 \\
NGC\,1155     &\nodata       &\nodata & &\nodata &\nodata &\nodata  & &CnB     &1690    &AS\,713 \\
UGC\,2982     &\nodata       &\nodata & &\nodata &\nodata &\nodata  & &C       &1650    &AS\,713 \\
NGC\,1569     &O63X09HYQ     &1320    & &HST$^b$ &1600    &F\,656N  & &B       &3910    &AA\,116 \\
NGC\,1614     &\nodata       &\nodata & &\nodata &\nodata &\nodata  & &C       &1710    &AS\,713 \\
              &\nodata       &\nodata & &\nodata &\nodata &\nodata  & &C       &1020    &AK\,331 \\
NGC\,1667     &O63X36ITQ     &1320    & &WIYN-2  &720     &KP\,1494 & &C       &1670    &AS\,713 \\
NGC\,1672     &O63X11R0Q     &1320    & &\nodata &\nodata &\nodata  & &\nodata &\nodata &\nodata \\
NGC\,1741     &O63X38FIQ     &1320    & &WIYN-2  &480     &KP\,1494 & &C       &990     &AK\,331 \\
              &\nodata       &\nodata & &\nodata &\nodata &\nodata  & &B       &1770$^d$&AM\,290 \\
NGC\,3079     &O63X35ZXQ     &1320    & &\nodata &\nodata &\nodata  & &CnB     &3050    &TT\,1   \\
NGC\,3690     &\nodata       &\nodata & &\nodata &\nodata &\nodata  & &C       &3740    &AS\,568 \\
NGC\,4088     &O63X12FMQ     &1320    & &WIYN-1  &480     &W\,015   & &C       &5380    &AL\,383 \\
NGC\,4100     &\nodata       &\nodata & &\nodata &\nodata &\nodata  & &C       &1680    &AS\,713 \\
NGC\,4214     &O63X39ESQ     &1320    & &HST$^c$ &1600    &F\,656N  & &C       &590     &AS\,713 \\
NGC\,4861     &O63X40H7Q     &1320    & &\nodata &\nodata &\nodata  & &C       &1670    &AS\,713 \\
NGC\,5054     &\nodata       &\nodata & &\nodata &\nodata &\nodata  & &C       &1700    &AS\,713 \\
NGC\,5161     &\nodata       &\nodata & &CTIO    &3600    &6600/75  & &C       &1690    &AS\,713 \\
NGC\,5383     &O63X16UKQ     &1320    & &\nodata &\nodata &\nodata  & &C       &1710    &AS\,713 \\
MRK\,799      &O63X17DRQ     &1320    & &WIYN-1  &720     &W\,016   & &C       &560     &AS\,713 \\
NGC\,5669     &O63X18AVQ     &1320    & &\nodata &\nodata &\nodata  & &C       &1700    &AS\,713 \\
NGC\,5676     &O63X19LSQ     &1320    & &\nodata &\nodata &\nodata  & &C       &1680    &AS\,713 \\
NGC\,5713     &O63X20FBQ     &1320    & &\nodata &\nodata &\nodata  & &C       &1690    &AS\,713 \\
NGC\,5860     &O63X21OUQ     &1320    & &WIYN-1  &720     &KP\,1495 & &C       &1690    &AS\,713 \\
NGC\,6090     &\nodata       &\nodata & &WIYN-1  &720     &KP\,1496 & &CnB     &1650    &AS\,713 \\
NGC\,6217     &O63X23BUQ     &1320    & &APO     &1200    &6590/25  & &CnB     &1660    &AS\,713 \\
NGC\,6643     &O63X24IAQ     &1320    & &APO     &1200    &6590/25  & &CnB     &1680    &AS\,713 \\
UGC\,11284    &O63X25Y6Q     &1320    & &\nodata &\nodata &\nodata  & &CnB     &1680    &AS\,713 \\
NGC\,6753     &\nodata       &\nodata & &CTIO    &4500    &6600/75  & &\nodata &\nodata &\nodata \\
TOL\,1924-416 &\nodata       &\nodata & &CTIO    &3600    &6600/75  & &CnB     &1690    &AS\,713 \\
NGC\,6810     &\nodata       &\nodata & &CTIO    &3600    &6600/75  & &\nodata &\nodata &\nodata \\
ESO\,400-G\,43&O63X29FKQ     &1320    & &CTIO    &3600    &6680/100 & &CnB     &570     &AS\,713 \\
              &\nodata       &\nodata & &\nodata &\nodata &\nodata  & &BnA     &3160$^d$&AJ\,176 \\
NGC\,7496     &\nodata       &\nodata & &CTIO    &3600    &6600/75  & &CnB     &1640    &AS\,713 \\
NGC\,7552     &\nodata       &\nodata & &CTIO    &3600    &6600/75  & &CnB     &1650    &AS\,713 \\
              &\nodata       &\nodata & &\nodata &\nodata &\nodata  & &A       &1590$^e$&AS\,721 \\
MRK\,323      &O63X31GSQ     &1320    & &WIYN-2  &960     &KP\,1494 & &C       &1680    &AS\,713 \\
NGC\,7673     &\nodata       &\nodata & &\nodata &\nodata &\nodata  & &C       &1670    &AS\,713 \\
NGC\,7714     &\nodata       &\nodata & &\nodata &\nodata &\nodata  & &C       &1680    &AS\,713 \\
MRK\,332      &O63X34AHQ     &1320    & &WIYN-2  &720     &W\,016   & &C       &1670    &AS\,713 \\
\enddata
\tablenotetext{.}{Column 1: galaxy name;
columns 2 and 3: the ultraviolet HST dataset name and exposure time in seconds,
respectively; columns 4, 5 and 6: the observatory where the H$\alpha$ images
were obtained (WIYN-1 and WIYN-2 correspond to the May 2001 and November 2001
observing runs), the exposure time of the images in seconds, and the name of
the filter used for the line observations; columns 7, 8 and 9: the VLA
configuration in which the galaxies were observed, the exposure times in
seconds and the code of the proposal from which the data were obtained.}
\tablenotetext{a}{Ultraviolet observations of IC\,1623 consist
of 3 exposures (O5CU02010, O5CU02020 and O5CU02030) obtained
as part of the HST project 8201 (P.I. Meurer).}
\tablenotetext{b}{Archival HST data from project 8133 (P.I. Shopbell).}
\tablenotetext{c}{Archival HST data from project 6569 (P.I. MacKenty).}
\tablenotetext{d}{These observations were done at 4.89~GHz.}
\tablenotetext{e}{This observation was done at 1.49~GHz.}
\label{tabobs}
\end{deluxetable}

\begin{deluxetable}{lrrrrrrrrrr}
\tabletypesize{\scriptsize}
\tablewidth{0pc}
\tablecaption{Ultraviolet and H$\alpha$ Fluxes}
\tablehead{\colhead{Name}&
\colhead{F(1457\AA)}&
\colhead{3$\sigma$}&
\colhead{F(H$\alpha$)$_{int}$}&
\colhead{F(H$\alpha$)$_{int}^{cor}$}&
\colhead{F(H$\alpha$)$_{match}$}&
\colhead{F(H$\alpha$)$_{match}^{cor}$}&
\colhead{3$\sigma$}&
\colhead{[NII]/H$\alpha$}&
\colhead{Area}&
\colhead{Ref.}\\
\colhead{(1)}&
\colhead{(2)}&
\colhead{(3)}&
\colhead{(4)}&
\colhead{(5)}&
\colhead{(6)}&
\colhead{(7)}&
\colhead{(8)}&
\colhead{(9)}&
\colhead{(10)}&
\colhead{(11)}}
\startdata
ESO\,350-G\,38& \nodata      &\nodata & 33.26$\pm$3.33&27.11$\pm$2.71&\nodata      &\nodata      & 7.1    & 0.17    &  1.5&   f \\
NGC\,232      & \nodata      &\nodata & 4.08$\pm$0.41 & 2.08$\pm$0.21&\nodata      &\nodata      & 4.7    & 0.72    &  9.2&   j \\
MRK\,555      & 1.43         &  2.8   & 9.06$\pm$0.91 & 6.66$\pm$0.67&3.72$\pm$0.37&2.73$\pm$0.27& 5.1    & 0.27    &\nodata&   k \\
IC\,1586      & \nodata      &\nodata & 2.29$\pm$0.22 & 2.29$\pm$0.22&\nodata      &\nodata      &\nodata & 0.21    &143.1&   a \\
NGC\,337      & \nodata      &\nodata & 32.09$\pm$3.21&24.32$\pm$2.43&\nodata      &\nodata      & 12.0   & 0.24    & 16.0&   l \\
IC\,1623      & 6.58         &  1.6   & \nodata       & \nodata      &\nodata      &\nodata      &\nodata & 0.30    &  5.2&   m \\
NGC\,1569     & 1.73         &  7.2   &151.85$\pm$15.20&151.09$\pm$15.10&136.00$\pm$13.60&135.30$\pm$13.50& 50.0   & 0.04    &  8.0&   h \\
NGC\,1614     & \nodata      &\nodata &10.69$\pm$1.07 &10.69$\pm$1.07&\nodata      &\nodata      &\nodata & 0.44    &143.1&   a \\
NGC\,1667     & 1.07         &  1.7   & 9.18$\pm$0.92 & 5.14$\pm$0.51&4.58$\pm$0.46&2.47$\pm$0.25& 9.7    & 0.69    &200.0&   b \\
NGC\,1672     & 3.89         &  2.5   &17.70$\pm$1.77 &17.70$\pm$1.77&17.70$\pm$1.77&17.70$\pm$1.77&\nodata & 0.46    &200.0&   b \\
NGC\,1741     & 3.95         &  3.7   & 12.17$\pm$1.22&10.73$\pm$1.07&10.20$\pm$1.02&9.00$\pm$0.90& 8.0    & 0.10    &\nodata&   i \\
NGC\,3079     & 0.46         & 19.3   & \nodata       & \nodata      &\nodata      &\nodata      &\nodata & 1.59    &  8.0&   h \\
NGC\,3690     & \nodata      &\nodata & 89.13$\pm$8.90&89.13$\pm$8.90&\nodata      &\nodata      &\nodata & 0.40    &  8.0&   c,h \\
NGC\,4088     & 0.67         & 13.5   & 62.14$\pm$6.21&47.85$\pm$4.79&4.36$\pm$0.44&3.36$\pm$0.34& 2.2    & 0.32    &  8.0&   h \\
NGC\,4214     & 18.36        &  5.7   & 83.16$\pm$8.32&81.66$\pm$8.17&76.80$\pm$7.68&75.02$\pm$7.50& 4.0    & 0.07    &  8.0&   h \\
NGC\,4861     & 13.07        &  2.4   & 19.75$\pm$1.98&19.75$\pm$1.98&19.75$\pm$1.98&19.75$\pm$1.98&\nodata & 0.09    &143.1&   a \\
NGC\,5161     & \nodata      &\nodata & 9.21$\pm$0.92 &7.09$\pm$0.71 &\nodata      &\nodata      & 17.3   & 0.36    & 19.6&   l \\
NGC\,5383     & 1.08         &  1.6   & 43.70$\pm$3.60&43.70$\pm$3.60&24.20$\pm$0.90&24.20$\pm$0.90&\nodata & 0.36    &  8.0&   d,h \\
MRK\,799      & 0.57         &  7.9   & 12.80$\pm$1.28& 8.44$\pm$0.84&4.28$\pm$0.43&2.82$\pm$0.28& 4.0    & 0.50    & 19.6&   j \\
NGC\,5669     & 1.09         & 13.4   & \nodata       & \nodata      &\nodata      &\nodata      &\nodata & 0.28    &  8.0&   h \\
NGC\,5676     & 0.37         &  8.9   & 23.56$\pm$2.30&23.56$\pm$2.30&\nodata      &\nodata      &\nodata & 0.45    &  8.0&   e,h \\
NGC\,5713     & 1.53         & 28.2   & \nodata       & \nodata      &\nodata      &\nodata      &\nodata & 0.47    & 16.6&   n \\
NGC\,5860     & 1.01         & 26.9   & 3.02$\pm$0.30 & 1.83$\pm$0.18&3.02$\pm$0.30&1.83$\pm$0.18& 2.1    & 0.49    &143.1&   a \\
NGC\,6090     & \nodata      &\nodata & 8.67$\pm$0.87 & 6.29$\pm$0.63&\nodata      &\nodata      & 3.3    & 0.45    &143.1&   a \\
NGC\,6217     & 1.86         &  3.0   & 20.45$\pm$2.05&18.47$\pm$1.85&7.99$\pm$0.80&7.21$\pm$0.72& 2.5    & 0.64    &143.1&   a \\
NGC\,6643     & 0.80         & 47.4   & 29.58$\pm$2.96&28.07$\pm$2.81&4.37$\pm$0.44&4.15$\pm$0.42& 6.4    & 0.32    &  8.0&   h \\
UGC\,11284    & 1.07         & 44.3   & \nodata       & \nodata      &\nodata      &\nodata      &\nodata & 0.38    &  7.2&   j \\
NGC\,6753     & \nodata      &\nodata & 22.60$\pm$2.26&22.60$\pm$2.26&\nodata      &\nodata      &\nodata & \nodata &\nodata&\nodata \\
TOL\,1924-416 & \nodata      &\nodata & 19.05$\pm$1.91&18.82$\pm$1.88&\nodata      &\nodata      & 6.5    & 0.02    &200.0&   b \\
NGC\,6810     & \nodata      &\nodata & 18.36$\pm$1.84&11.86$\pm$1.19&\nodata      &\nodata      & 5.9    & 0.53    &  8.0&   l \\
ESO\,400-G\,43& 3.57         & 43.0   & 14.25$\pm$1.43&13.35$\pm$1.34&14.10$\pm$1.41&13.21$\pm$1.30& 11.2   & 0.05    &\nodata&   g \\
NGC\,7496     & \nodata      &\nodata & 33.00$\pm$3.30&20.72$\pm$2.07&\nodata      &\nodata      & 17.5   & 0.48    &200.0&   b \\
NGC\,7552     & \nodata      &\nodata & 71.00$\pm$7.10&41.61$\pm$4.16&\nodata      &\nodata      & 15.6   & 0.57    &200.0&   b \\
MRK\,323      & 0.79         & 34.3   & 2.88$\pm$0.29 & 2.88$\pm$0.29&2.13$\pm$0.21&2.13$\pm$0.21& 6.6    & \nodata &\nodata&\nodata \\
NGC\,7673     & \nodata      &\nodata & 6.08$\pm$0.60 & 6.08$\pm$0.60&\nodata      &\nodata      &\nodata & 0.23    &143.1&   a \\
NGC\,7714     & \nodata      &\nodata & 27.96$\pm$2.80&27.96$\pm$2.80&\nodata      &\nodata      &\nodata & 0.36    &143.1&   a \\
MRK\,332      & 1.37         &  5.2   & 13.58$\pm$1.36& 7.33$\pm$0.73&9.22$\pm$0.92&4.98$\pm$0.50& 4.4    & 0.64    &  8.0&   h \\
\enddata
\tablenotetext{a}{
Column 1: galaxy name;
column 2: ultraviolet flux in units of 10$^{-14}$~erg~cm$^{-2}$~s$^{-1}$~\AA$^{-1}$, not corrected for Galactic extinction, the accuracy of the fluxes is of the order of 5\%;
column 3: the 3$\sigma$ detection limit of the UV images in units of
10$^{-20}$~erg~cm$^{-2}$~s$^{-1}$~\AA$^{-1}$~pix$^{-1}$;
column 4: integrated H$\alpha$ flux, not corrected for
[NII]$\lambda$6548,84\AA\ contamination, in units of 
10$^{-13}$~erg~cm$^{-2}$~s$^{-1}$.
column 5: integrated H$\alpha$ flux, corrected for [NII] contamination.
Except for NGC\,6217 and MRK\,323, galaxies with identical values in
columns 4 and 5 correspond to values obtained from the literature;
column 6: H$\alpha$ flux measured inside an aperture matching that of the UV
image, not corrected for [NII]$\lambda$6548,84\AA\ contamination, in units of
10$^{-13}$~erg~cm$^{-2}$~s$^{-1}$. Notice that
the values for NGC\,1672 and NGC\,5383 were obtained from the literature
and are not contaminated by [NII];
column 7: Same as column 6, but corrected for [NII] contamination;
column 8: the 3$\sigma$ detection limit of the H$\alpha$ images,
in units of 10$^{-18}$~erg~cm$^{-2}$~s$^{-1}$~pix$^{-1}$;
column 9: [NII]/H$\alpha$ emission line ratio; column 10: the area
of the slit used to observe the [NII]/H$\alpha$ ratio, In the case of
ESO\,400--G\,43 the observations were centered at the nucleus, but the observed
area was not given in the paper. NGC\,1741 was observed with a slit 1.5\arcsec\
wide, but the length of the extraction was not informed. MRK\,555 was observed
with a slit narrower than 8\arcsec, but the size of the extraction was not informed;
column 11: references from which we obtained the emission line ratio [NII]/H$\alpha$,
and for those galaxies for which we did not obtain images, the H$\alpha$ flux.
a-) McQuade, Calzetti \& Kinney (1995);
b-) Storchi-Bergmann, T., Kinney, A. L. \& Challis, P. (1995);
c-) Armus, Heckman \& Miley (1990);
d-) Sheth et al. (2000);
e-) Kennicutt \& Kent (1983);
f-) Kewley et al. (2000);
g-) Fairall (1988);
h-) Ho, Filippenko \& Sargent (1997);
i-) Vacca \& Conti (1992);
j-) Veilleux et al. (1995);
k-) Terlevich et al. (1991);
l-) Veron-Cetty \& Veron (1986);
m-) Corbett et al. (2003);
n-) Kewley et al. (2001).}
\label{tabuvha}
\end{deluxetable}

\begin{deluxetable}{lrrrrrrr}
\tabletypesize{\scriptsize}
\tablewidth{0pc}
\tablecaption{Radio Fluxes}
\tablehead{\colhead{Name}&
\colhead{S(1.49~GHz)}&
\colhead{S(4.89~GHz)}&
\colhead{S(8.46~GHz)$_{int}$}&
\colhead{S(8.46~GHz)$_{match}$}&
\colhead{3$\sigma$}&
\colhead{Beam}&
\colhead{Ref.}\\
\colhead{}&
\colhead{(mJy)}&
\colhead{(mJy)}&
\colhead{(mJy)}&
\colhead{(mJy)}&
\colhead{($\mu$Jy)}&
\colhead{(arcsec)}&
\colhead{}\\
\colhead{(1)}&
\colhead{(2)}&
\colhead{(3)}&
\colhead{(4)}&
\colhead{(5)}&
\colhead{(6)}&
\colhead{(7)}&
\colhead{(8)}}
\startdata
ESO\,350-G\,38&  27.2$\pm$0.9 &   15.1$\pm$0.5&   10.1$\pm$0.2 &\nodata          & 61 &1.37$\times$0.65 & a \\
NGC\,232      &  60.6$\pm$1.9 &     56$\pm$11 &   14.9$\pm$0.3 &\nodata          & 70 &3.69$\times$2.45 & h \\
MRK\,555      &  39.6$\pm$1.9 &     10$\pm$4  &    5.2$\pm$0.2 &    3.1$\pm$0.1 & 61 &4.09$\times$2.96 & b \\
IC\,1586      &   8.3$\pm$0.5 &\nodata        &    1.7$\pm$0.1 &\nodata          & 83 &4.11$\times$2.86 & \nodata \\
NGC\,337      & 109.6$\pm$4.1 &     44$\pm$11 &    9.1$\pm$0.3 &\nodata          & 56 &5.08$\times$2.99 & i \\
IC\,1623      & 249.2$\pm$9.8 &     96$\pm$12 &   49.6$\pm$0.8 &   49.0$\pm$0.8 & 93 &3.34$\times$2.24 & h \\
NGC\,1155     &   9.1$\pm$0.6 &\nodata        &    2.1$\pm$0.1 &\nodata          & 83 &3.14$\times$2.19 & \nodata \\
UGC\,2982     &  92.4$\pm$3.7 &     28$\pm$1  &   18.2$\pm$0.4 &\nodata          & 62 &3.57$\times$2.69 & c \\
NGC\,1569     & 338.6$\pm$11.0&    202$\pm$19 &   23.2$\pm$0.5 &   17.0$\pm$0.4 & 64 &1.06$\times$0.84 & j \\
NGC\,1614     & 138.2$\pm$4.9 &     63$\pm$11 &   41.1$\pm$0.6 &\nodata          & 71 &3.72$\times$2.58 & i \\
NGC\,1667     &  77.3$\pm$3.0 &     45$\pm$11 &   17.7$\pm$0.4 &    7.5$\pm$0.2 & 92 &4.23$\times$2.78 & i \\
NGC\,1672     & 450.0$\pm$45.0&    114$\pm$9  &\nodata          &\nodata          &\nodata&\nodata& k \\
NGC\,1741     &  31.7$\pm$1.6 &    6.9$\pm$0.6&    5.8$\pm$0.2 &    5.0$\pm$0.2 &105 &3.50$\times$2.58 & a \\
NGC\,3079     & 770.7$\pm$27.1&    321$\pm$34 &  120.2$\pm$1.9 &  109.9$\pm$1.7 &107 &1.92$\times$0.99 & l \\
NGC\,3690     & 678.1$\pm$25.4&    300$\pm$4  &  226.0$\pm$3.4 &\nodata          & 79 &3.34$\times$2.58 & c \\
NGC\,4088     & 174.1$\pm$5.9 &     67$\pm$9  &   22.2$\pm$0.5 &    3.9$\pm$0.1 & 80 &3.00$\times$2.71 & j \\
NGC\,4100     &  50.3$\pm$2.2 &\nodata        &   26.2$\pm$0.5 &\nodata          &105 &3.99$\times$2.94 & \nodata \\
NGC\,4214     &  34.8$\pm$1.5 &     30$\pm$7  &   20.5$\pm$0.5 &    9.7$\pm$0.2 &112 &3.75$\times$3.05 & k \\
NGC\,4861     &  14.4$\pm$0.9 &    8.3$\pm$0.4&    6.3$\pm$0.3 &    6.3$\pm$0.3 & 79 &4.19$\times$3.17 & d \\
NGC\,5054     &  89.7$\pm$3.4 &     49$\pm$11 &   17.3$\pm$0.4 &\nodata          & 67 &5.33$\times$2.63 & h \\
NGC\,5161     &  11.3$\pm$2.5 &\nodata        &    1.1$\pm$0.2 &\nodata          & 60 &8.38$\times$2.62 & \nodata \\
NGC\,5383     &  30.7$\pm$1.6 &     11$\pm$5  &    5.9$\pm$0.2 &    5.4$\pm$0.2 & 88 &4.42$\times$3.02 & b \\
MRK\,799      &  67.0$\pm$2.7 &     29$\pm$6  &   14.3$\pm$0.4 &    9.1$\pm$0.2 & 96 &4.25$\times$2.76 & j \\
NGC\,5669     &  19.0$\pm$2.7 &\nodata        &\nodata          &\nodata          & 90 &4.68$\times$2.95 & \nodata \\
NGC\,5676     & 118.8$\pm$4.2 &     38$\pm$6  &   15.4$\pm$0.5 &    4.6$\pm$0.2 & 76 &4.16$\times$2.90 & j \\
NGC\,5713     & 159.9$\pm$5.7 &     93$\pm$14 &   30.6$\pm$0.6 &   19.2$\pm$0.4 & 81 &3.92$\times$2.69 & l \\
NGC\,5860     &   7.7$\pm$0.5 &\nodata        &    1.9$\pm$0.3 &    1.9$\pm$0.3 &115 &3.19$\times$2.75 & \nodata \\
NGC\,6090     &  48.4$\pm$1.5 &   20.3$\pm$1.6&   12.9$\pm$0.3 &\nodata          & 75 &2.80$\times$1.58 & e \\
NGC\,6217     &  80.8$\pm$3.1 &     21$\pm$1  &   13.2$\pm$0.2 &   13.1$\pm$0.2 & 81 &3.26$\times$2.13 & f \\
NGC\,6643     &  97.8$\pm$3.6 &     34$\pm$5  &    1.8$\pm$0.1 &    0.4$\pm$0.1 & 75 &2.98$\times$2.21 & j \\
UGC\,11284    &  60.8$\pm$2.5 &\nodata        &   13.0$\pm$0.3 &    5.5$\pm$0.1 & 70 &2.66$\times$2.50 & \nodata \\
NGC\,6753     &\nodata        &     35$\pm$8  &\nodata          &\nodata          &\nodata&\nodata& k \\
TOL\,1924-416 &\nodata        &\nodata        &    6.2$\pm$0.2 &\nodata          & 86 &4.90$\times$2.56 & \nodata \\
NGC\,6810     &\nodata        &     72$\pm$8  &\nodata          &\nodata          &\nodata&\nodata& k \\
ESO\,400-G\,43&  18.1$\pm$1.0 &    5.3$\pm$0.5&    4.3$\pm$0.2 &    4.1$\pm$0.2 &150 &5.36$\times$2.03 & a \\
NGC\,7496     &  35.0$\pm$1.8 &\nodata        &    7.9$\pm$0.2 &\nodata          & 87 &5.48$\times$2.57 & \nodata \\
NGC\,7552     & 216.8$\pm$8.4 &    139$\pm$11 &   66.9$\pm$1.1 &\nodata          & 90 &5.04$\times$2.58 & k \\
MRK\,323      &  25.7$\pm$1.4 &\nodata        &    4.3$\pm$0.2 &    3.3$\pm$0.2 & 80 &2.94$\times$2.76 & \nodata \\
NGC\,7673     &  32.0$\pm$1.9 &     17$\pm$2  &   10.0$\pm$0.2 &\nodata          & 67 &2.99$\times$2.87 & g \\
NGC\,7714     &  66.9$\pm$2.7 &     39$\pm$9  &   19.1$\pm$0.4 &\nodata          &102 &4.30$\times$2.87 & j \\
MRK\,332      &  37.5$\pm$1.7 &\nodata        &    6.0$\pm$0.3 &    4.0$\pm$0.2 &105 &3.03$\times$2.89 & \nodata \\
\enddata
\tablenotetext{.}{
Column 1: galaxy name;
column 2: 1.49~GHz (20cm) fluxes obtained from the NVSS (Condon et al. 1998),
except for NGC\,7496, which was obtained from Condon (1987), and NGC\,7552,
which was obtained from our own data;
column 3: 4.89~GHz (6 cm) fluxes, obtained from the  literature, or from new data;
column 4: integrated 8.46~GHz (3.5cm) fluxes, obtained from the data presented
in this paper;
column 5: 8.46~GHz fluxes measured inside an aperture matching the ultraviolet one;
column 6: the 3$\sigma$ detection limit of the 8.46~GHz images;
column 7: the beam of the 8.46~GHz images;
column 8: the references from which the 4.89~GHz fluxes were obtained:
a-) This paper;
b-) Bicay et al. (1995);
c-) Condon, Anderson \& Broderick (1995);
d-) Klein, Wielebinski \& Thuan (1984);
e-) van der Hulst, Crane \& Keel (1981);
f-) Saikia et al. (1994);
g-) Condon \& Yin (1990);
h-) Griffith et al. (1994);
i-) Griffith et al. (1995);
j-) Gregory \& Condon (1991);
k-) Wright et al. (1994b);
l-) Becker, White \& Edwards (1991).}
\label{tabrad}
\end{deluxetable}

\begin{deluxetable}{lrrrrrrrrr}
\tabletypesize{\scriptsize}
\tablewidth{0pc}
\tablecaption{Infrared Fluxes}
\tablehead{\colhead{Name}&
\colhead{12$\mu$m}&
\colhead{25$\mu$m}&
\colhead{60$\mu$m}&
\colhead{100$\mu$m}&
\colhead{150$\mu$m}&
\colhead{170$\mu$m}&
\colhead{180$\mu$m}&
\colhead{205$\mu$m}&
\colhead{Ref.}\\
\colhead{}&
\colhead{(Jy)}&
\colhead{(Jy)}&
\colhead{(Jy)}&
\colhead{(Jy)}&
\colhead{(Jy)}&
\colhead{(Jy)}&
\colhead{(Jy)}&
\colhead{(Jy)}&
\colhead{}\\
\colhead{(1)}&
\colhead{(2)}&
\colhead{(3)}&
\colhead{(4)}&
\colhead{(5)}&
\colhead{(6)}&
\colhead{(7)}&
\colhead{(8)}&
\colhead{(9)}&
\colhead{(10)}}
\startdata
ESO\,350-G\,38&  0.42&  2.49&  6.48&  5.01& \nodata&  \nodata& \nodata& \nodata&\nodata  \\
NGC\,232      &  0.33&  1.08& 10.04& 18.34&    18.9$\pm$5.2&     15.7$\pm$3.9&    9.1$\pm$1.5&    5.4$\pm$0.1&a  \\
MRK\,555      &  0.28&  0.56&  4.22&  8.68& \nodata&  \nodata& \nodata&     5.3$\pm$1.6&b  \\
IC\,1586    &$<$0.12&$<$0.21&  0.96&  1.69&    2.1$\pm$0.5&  \nodata& \nodata&    0.8$\pm$0.4&c  \\
NGC\,337      &  0.22&  0.65&  8.35& 17.11& \nodata&  \nodata& \nodata& \nodata&\nodata  \\
IC\,1623      &  0.68&  3.57& 22.58& 30.37&    26.7$\pm$0.3&     23.5$\pm$0.4&    12.8$\pm$0.1&    9.7$\pm$0.1&a  \\
NGC\,1155     &  0.17&  0.36&  2.45&  4.60& \nodata&  \nodata& \nodata& \nodata&\nodata  \\
UGC\,2982     &  0.55&  0.78&  8.35& 16.89&    12.5$\pm$3.8&      9.8$\pm$2.9&     9.9$\pm$3.0&     9.2$\pm$2.8&b  \\
NGC\,1569     &  0.79&  7.09& 45.41& 47.29& \nodata&  \nodata& \nodata& \nodata&\nodata  \\
NGC\,1614     &  1.44&  7.29& 32.31& 32.69& \nodata&    17.1$\pm$0.7& \nodata& \nodata&d  \\
NGC\,1667     &  0.43&  0.68&  5.95& 14.73&    16.3$\pm$0.5&     17.0$\pm$2.8&    9.0$\pm$0.2&    6.3$\pm$0.2&a  \\
NGC\,1672     &  1.67&  4.03& 32.96& 69.89& \nodata&  \nodata& \nodata& \nodata&\nodata  \\
NGC\,1741     &  0.11&  0.58&  3.92&  5.84& \nodata&  \nodata& \nodata& \nodata&\nodata  \\
NGC\,3079     &  1.52&  2.27& 44.50& 89.22&   125.7$\pm$18.9&  \nodata&   136.9$\pm$20.6&   134.4$\pm$20.3&e  \\
NGC\,3690     &  3.81& 23.19&103.70&107.40& \nodata&  \nodata& \nodata& \nodata&\nodata  \\
NGC\,4088     &  0.88&  1.55& 19.88& 54.47& \nodata&    144.8$\pm$2.5&    46.2$\pm$9.2& \nodata&d,f  \\
NGC\,4100     &  0.50&  0.82&  8.10& 21.72& \nodata&  \nodata&    18.6$\pm$3.7& \nodata&f  \\
NGC\,4214     &  0.61&  2.36& 17.87& 29.04& \nodata&  \nodata& \nodata& \nodata&\nodata  \\
NGC\,4861    &$<$0.13&  0.39&  1.97&  2.46& \nodata&  \nodata& \nodata& \nodata&\nodata  \\
NGC\,5054     &  0.76&  1.15& 11.60& 26.21& \nodata&  \nodata&    26.7$\pm$5.3& \nodata&f  \\
NGC\,5161     &  0.18&  0.25&  2.18&  7.24& \nodata&  \nodata& \nodata& \nodata&\nodata  \\
NGC\,5383     &  0.35&  0.69&  4.89& 13.70& \nodata&  \nodata& \nodata& \nodata&\nodata  \\
MRK\,799      &  0.56&  1.63& 10.41& 19.47&    12.9$\pm$3.9&  \nodata&    11.4$\pm$3.4&     9.3$\pm$2.8&b  \\
NGC\,5669     &  0.09&  0.12&  1.66&  5.19& \nodata&  \nodata&    6.8$\pm$1.4& \nodata&f  \\
NGC\,5676     &  0.71&  1.03&  9.64& 30.66& \nodata&  \nodata&    25.8$\pm$5.2& \nodata&f  \\
NGC\,5713     &  1.10&  2.58& 19.82& 36.20& \nodata&  \nodata&    22.3$\pm$4.5& \nodata&f  \\
NGC\,5860     &  0.13&  0.20&  1.64&  3.02&    2.6$\pm$0.7&  \nodata& \nodata&    0.7$\pm$0.3&c  \\
NGC\,6090     &  0.26&  1.11&  6.66&  8.94&    8.7$\pm$2.2&  \nodata& \nodata&    4.5$\pm$1.1&c  \\
NGC\,6217     &  0.51&  1.61& 10.83& 19.33& \nodata&    26.2$\pm$0.6&    16.1$\pm$3.2& \nodata&d,f  \\
NGC\,6643     &  0.81&  1.04&  9.38& 30.69& \nodata&  \nodata&    24.7$\pm$4.9& \nodata&f  \\
UGC\,11284    &  0.36&  1.03&  8.25& 15.18&    27.0$\pm$0.2&  \nodata&    14.5$\pm$0.2&    13.5$\pm$0.3&a  \\
NGC\,6753     &  0.60&  0.73&  9.43& 27.36& \nodata&  \nodata&    19.0$\pm$3.8& \nodata&f  \\
TOL\,1924-416&$<$0.07&  0.42&  1.69&  1.01&    0.7$\pm$0.2&  \nodata& \nodata&    0.1$\pm$0.1&c  \\
NGC\,6810     &  1.10&  3.49& 17.79& 34.50&    29.8$\pm$0.1&  \nodata&    16.9$\pm$0.2&    14.0$\pm$0.2&a  \\
ESO\,400-G\,43&  0.10&  0.21&  1.59&  1.58& \nodata&  \nodata& \nodata& \nodata&\nodata  \\
NGC\,7496     &  0.35&  1.60&  8.46& 15.55&    16.2$\pm$0.2&     13.9$\pm$0.5&    8.9$\pm$0.1&    6.3$\pm$0.1&a  \\
NGC\,7552     &  2.95& 12.16& 72.03&101.50& \nodata&    104.7$\pm$3.6& \nodata& \nodata&d  \\
MRK\,323      &  0.27&  0.35&  3.16&  7.91&     7.0$\pm$2.1&  \nodata&     6.5$\pm$2.0&     6.0$\pm$1.8&b  \\
NGC\,7673     &  0.13&  0.52&  4.91&  6.89&     7.6$\pm$1.9&  \nodata& \nodata&    4.1$\pm$1.0&c  \\
NGC\,7714     &  0.47&  2.85& 10.36& 11.51&     8.0$\pm$1.1&  \nodata&     5.3$\pm$0.8&     5.5$\pm$0.8&g  \\
MRK\,332      &  0.36&  0.62&  4.87&  9.49&     6.4$\pm$1.9&  \nodata&     5.6$\pm$1.7&     4.7$\pm$1.4&b  \\
\enddata
\tablenotetext{.}{Column 1: galaxy name;
columns 2, 3, 4 and 5: the IRAS 12, 25, 60 and 100$\mu$m fluxes, respectively,
we assume that the error of these fluxes is 6\%;
columns 6, 7, 8 and 9: the ISO 150, 170, 180 and 205 $\mu$m fluxes, respectively;
column 9:references from which the ISO fluxes were obtained. In the cases where
more than one reference is given, the first one corresponds to the 170$\mu$m flux
and the second one to 180$\mu$m. a-) Spinoglio, Andreani \& Malkan (2002);
b-) Siebenmorgen, Krugel \& Chini (1999); c-) Calzetti et al. (2000);
d-) Stickel et al. (2000); e-) Perez Garcia, Rodriguez Espinosa \& 
Santolaya Rey (1998); f-) Bendo et al. (2002); g-) Krugel et al. (1998).}
\label{tabir}
\end{deluxetable}

\begin{deluxetable}{lrrrrrrrrrrrrr}
\tabletypesize{\scriptsize}
\tablewidth{0pc}
\tablecaption{Ultraviolet, Optical and Near Infrared Fluxes}
\tablehead{\colhead{Name}&
\colhead{F(1482\AA)}&
\colhead{F(1913\AA)}&
\colhead{F(2373\AA)}&
\colhead{F(2700\AA)}&
\colhead{}&
\colhead{U}&
\colhead{B}&
\colhead{V}&
\colhead{R}&
\colhead{J}&
\colhead{H}&
\colhead{K}&
\colhead{Ref.}\\
\cline{2-5}\cline{7-13}\\
\colhead{}&
\multicolumn{4}{c}{(10$^{-14}$ erg cm$^{-2}$ s$^{-1}$ \AA$^{-1}$)}&
\colhead{}&
\multicolumn{7}{c}{(mJy)}&
\colhead{}\\
\colhead{(1)}&
\colhead{(2)}&
\colhead{(3)}&
\colhead{(4)}&
\colhead{(5)}&
\colhead{}&
\colhead{(6)}&
\colhead{(7)}&
\colhead{(8)}&
\colhead{(9)}&
\colhead{(10)}&
\colhead{(11)}&
\colhead{(12)}&
\colhead{(13)}}
\startdata
ESO\,350-G\,38& \nodata&\nodata&\nodata&\nodata& &\nodata&   8.0&\nodata&   12.4& 10.5& 11.8& 11.5& b\\
NGC\,232      & \nodata&\nodata&\nodata&\nodata& &\nodata&   7.2&\nodata&   23.9& 64.3& 78.7& 73.4& a\\
MRK\,555      & \nodata&\nodata&\nodata&\nodata& &\nodata&  27.6&\nodata&\nodata& 83.5&  104.0& 91.5& c\\
IC\,1586      &    0.59$\pm$0.14&   0.48$\pm$0.05&\nodata&\nodata& &\nodata&   4.7&   5.1&    7.8& 10.8& 13.2& 10.8& c\\
NGC\,337      & \nodata&\nodata&\nodata&\nodata& &  29.5&  63.9&   82.6&\nodata&  163.0&  180.0&  153.0& a\\
IC\,1623      & \nodata&\nodata&\nodata&\nodata& &\nodata&   6.2&\nodata&\nodata& 37.2& 55.4& 54.8& b\\
NGC\,1155     & \nodata&\nodata&\nodata&\nodata& &\nodata&   2.5&\nodata&\nodata& 32.0& 35.7& 31.0& c\\
UGC\,2982     & \nodata&\nodata&\nodata&\nodata& &\nodata&   3.3&\nodata&\nodata& 56.3& 74.0& 79.0& a\\
NGC\,1569     &    1.17$\pm$0.16&   0.81$\pm$0.12&   0.69$\pm$0.13&   1.33$\pm$0.07& &  37.1&   76.8&    141.0&    147.0&  471.0&  549.0&  480.0& a\\
NGC\,1614     &    0.56$\pm$0.22&   0.43$\pm$0.08&\nodata&\nodata& &   6.5&   15.1&   24.3&   23.6& 40.3& 46.6& 53.2& d\\
NGC\,1667     &    0.30$\pm$0.18&   0.26$\pm$0.09&\nodata&\nodata& &   13.7&   33.2&   54.1&\nodata&  171.0&  218.0&  184.0& c\\
NGC\,1672     &    3.01$\pm$0.31&   2.77$\pm$0.18&   2.85$\pm$0.67&   2.90$\pm$0.24& &    139.0&  329.0&  489.0&  473.0& 1100.0& 1190.0& 1040.0& a\\
NGC\,1741     & \nodata&\nodata&\nodata&\nodata& &   14.4&   20.4&   24.7&\nodata& 12.2& 13.9& 12.2& c\\
NGC\,3079     & \nodata&\nodata&\nodata&\nodata& &   42.6&  103.0&    165.0&\nodata&  670.0&  903.0&  830.0& a\\
NGC\,3690     &    2.33$\pm$0.27&   1.68$\pm$0.15&   1.08$\pm$0.31&   1.02$\pm$0.15& &\nodata&\nodata&   66.2&\nodata&  224.0&  296.0&  285.0& a\\
NGC\,4088     & \nodata&\nodata&\nodata&\nodata& &   65.7&    148.0&    217.0&\nodata&  649.0&  797.0&  680.0& a\\
NGC\,4100     & \nodata&\nodata&\nodata&\nodata& &   29.5&   74.7&    125.0&\nodata&  413.0&  501.0&  408.0& a\\
NGC\,4214     &   10.30$\pm$0.93&   6.34$\pm$0.39&   4.34$\pm$0.34&   3.55$\pm$0.12& &    141.0&    342.0&    446.0&\nodata&  520.0&  614.0&  458.0& a\\
NGC\,4861     &    9.92$\pm$0.98&   5.47$\pm$0.85&   3.34$\pm$0.86&   2.56$\pm$0.67& &\nodata&   29.5&   43.0&\nodata& 16.8& 14.0& 13.1& a\\
NGC\,5054     & \nodata&\nodata&\nodata&\nodata& &   32.9&   91.5&    157.0&    203.0&  622.0&  760.0&  614.0& a\\
NGC\,5161     & \nodata&\nodata&\nodata&\nodata& &   26.7&   67.5&    119.0&    134.0&  254.0&  296.0&  235.0& a\\
NGC\,5383     & \nodata&\nodata&\nodata&\nodata& &   25.7&   64.5&    100.0&\nodata&  249.0&  297.0&  257.0& a\\
MRK\,799      & \nodata&\nodata&\nodata&\nodata& &\nodata&   34.8&\nodata&\nodata&  157.0&  194.0&  170.0& c\\
NGC\,5669     & \nodata&\nodata&\nodata&\nodata& &\nodata&   65.7&\nodata&\nodata&  141.0&  247.0&  169.0& e\\
NGC\,5676     & \nodata&\nodata&\nodata&\nodata& &   30.3&   76.1&    122.0&\nodata&  410.0&  540.0&  456.0& c\\
NGC\,5713     & \nodata&\nodata&\nodata&\nodata& &   32.0&   78.2&    120.0&\nodata&  328.0&  369.0&  310.0& a\\
NGC\,5860     &    0.74$\pm$0.21&   0.50$\pm$0.11&   0.37$\pm$0.11&   0.46$\pm$0.08& &\nodata&    8.9&\nodata&\nodata& 35.6& 42.6& 32.5& c\\
NGC\,6090     &    1.07$\pm$0.34&   0.87$\pm$0.11&\nodata&\nodata& &\nodata&   10.7&    8.6&   14.2& 16.7& 21.4& 20.0& h\\
NGC\,6217     &    1.74$\pm$0.21&   1.55$\pm$0.15&\nodata&\nodata& &   41.1&   81.9&    125.0&\nodata&  214.0&  260.0&  226.0& c\\
NGC\,6643     & \nodata&\nodata&\nodata&\nodata& &   38.2&   86.6&    133.0&\nodata&  297.0&  383.0&  325.0& g\\
UGC\,11284    & \nodata&\nodata&\nodata&\nodata& &\nodata&    3.2&\nodata&\nodata& 11.7& 19.2& 19.1& b\\
NGC\,6753     & \nodata&\nodata&\nodata&\nodata& &   26.2&   69.4&    166.0&    231.0&  632.0&  780.0&  724.0& a\\
TOL\,1924-416 &    3.81$\pm$0.25&   2.25$\pm$0.33&   1.41$\pm$0.11&   1.19$\pm$0.05& &   12.8&   20.4&   22.8&   11.9& 7.1& 7.2& 5.9& c\\
NGC\,6810     & \nodata&\nodata&\nodata&\nodata& &   15.5&   48.0&   96.6&    161.0&  477.0&  656.0&  565.0& c\\
ESO\,400-G\,43& \nodata&\nodata&\nodata&\nodata& &\nodata&   8.4&\nodata&   8.2& 8.2& 10.7& 6.1& c\\
NGC\,7496     &    1.43$\pm$0.21&   1.06$\pm$0.08&   1.00$\pm$0.14&   0.87$\pm$0.05& &\nodata&   72.8&\nodata&    129.0&  255.0&  298.0&  231.0& c\\
NGC\,7552     &    1.72$\pm$0.28&   1.91$\pm$0.19&   1.94$\pm$0.56&   2.32$\pm$0.23& &   52.7&    138.0&  215.0&    285.0&  638.0&  749.0&  645.0& a\\
MRK\,323      & \nodata&\nodata&\nodata&\nodata& &   5.1&   12.6&   20.6&\nodata& 72.4& 87.8& 74.3& c\\
NGC\,7673     &    2.29$\pm$0.37&   1.52$\pm$0.11&   1.03$\pm$0.16&   1.00$\pm$0.07& &   13.2&   23.0&   28.7&\nodata& 39.6& 44.5& 34.0& c\\
NGC\,7714     &    3.70$\pm$0.41&   2.55$\pm$0.26&   1.88$\pm$0.27&   2.00$\pm$0.13& &   17.3&   26.9&   37.1&\nodata& 78.1& 89.5& 83.0& c\\
MRK\,332      & \nodata&\nodata&\nodata&\nodata& &   13.4&   27.6&   41.0&   63.1&  115.0&  141.0&  119.0& c\\
\enddata
\tablenotetext{.}{Column 1: Galaxy name;
columns 2, 3, 4 and 5: ultraviolet IUE fluxes from Kinney et al. (1993),
obtained inside an aperture of 10\arcsec $\times$20\arcsec, not corrected for
Galactic extinction;
columns 6, 7, 8 and 9: the U, B, V and R fluxes obtained from NED, giving preference
to the largest aperture measurements available, the uncertainty of these fluxes
is of the order of 10-15\%;
columns 10, 11 and 12: the near infrared J, H and K band fluxes, the uncertainty
of these fluxes is of the order of 5\%;
column 13: the references from which the near infrared data were obtained.
a-) Jarrett et al. (2003);
b- Spinoglio et al. (1995);
c- Ned 2MASS;
d- Balzano \& Weedman (1981);
e- Bendo et al. (2002);
f- Balzano (1983);
g- Aaronson (1977);
h- Wu et al. (2002)}
\label{tabanc}
\end{deluxetable}

\begin{deluxetable}{lrrr}
\tabletypesize{\scriptsize}
\tablewidth{0pc}
\tablecaption{Spectral Energy Distributions}
\tablehead{
\colhead{Band}&
\colhead{Log $\nu$}&
\colhead{Log $\nu F_{\nu}$ (REDSED)}&
\colhead{Log $\nu F_{\nu}$ (BLUESED)}\\
\colhead{}&
\colhead{(Hz)}&
\colhead{}&
\colhead{}}
\startdata
1457\AA   &  15.31 & 3.83 & 4.26 \\
B         &  14.83 & 4.89 & 5.58 \\
J         &  14.38 & 4.94 & 5.78 \\
H         &  14.26 & 4.92 & 5.76 \\
K         &  14.14 & 4.77 & 5.56 \\
12$\mu$m  & 13.40 & 4.95 & 5.13 \\
25$\mu$m  & 13.08 & 5.15 & 5.11 \\
60$\mu$m  & 12.70 & 5.61 & 5.65 \\
100$\mu$m & 12.48 & 5.53 & 5.78 \\
150$\mu$m & 12.30 & 5.34 & 5.51 \\
170$\mu$m & 12.25 & 5.21 & 5.42 \\
180$\mu$m & 12.22 & 5.06 & 5.39 \\
205$\mu$m & 12.16 & 4.87 & 5.09 \\
8.46~GHz  &  9.93 & 0.13 & 0.06 \\
4.89~GHz  &  9.69 & 0.17 & 0.09 \\
1.49~Ghz  &  9.15 & 0.00 & 0.00 \\
\enddata
\tablenotetext{.}{Columns 1 and 2 indicate the band and the corresponding
logarithm of the frequency, respectively; columns 3 and 4 give the average
SEDs of galaxies dominated by the far infrared emission (REDSED) and
those with similar contributions from the optical and the far infrared
part of the spectrum (BLUESED), respectively.
These SED are normalized to the energy density at 1.49~Ghz.}
\label{tabsed}
\end{deluxetable}

\end{document}